\documentclass[runningheads]{llncs}
\usepackage{graphicx}
\usepackage{comment}
\usepackage{amsmath,amssymb} 
\usepackage{float}
\usepackage{subcaption}
\usepackage{mathtools}
\usepackage{color}
\usepackage{multirow}
\usepackage{cite}

\newcommand{\red}[1]{{\color{red}{#1}}} 
\newcommand{\blue}[1]{{\color{blue}{#1}}} 
\newcommand{\etal}{et al.}

\begin{document}
\pagestyle{headings}
\mainmatter
\def\ECCVSubNumber{3269}  

\title{Exploring Multi-Scale Feature Propagation and Communication for Image Super Resolution} 


\titlerunning{Multi-Scale Feature Propagation and Communication}
%
\author{
Ruicheng Feng\inst{1} \and
Weipeng Guan\inst{1} \and
Yu Qiao\inst{1,2} \and
Chao Dong\inst{1}
}
\authorrunning{R. Feng et al.}
%
\institute{$^1$Shenzhen Key Lab of Computer Vision and Pattern Recognition, SIAT-SenseTime Joint Lab, Shenzhen Institutes of Advanced Technology, Chinese Acedamy of Sciences \\
$^2$The Chinese University of Hong Kong\\
\email{\{rc.feng, wp.guan, yu.qiao, chao.dong\}@siat.ac.cn}}
\maketitle

\begin{abstract}
Multi-scale techniques have achieved great success in a wide range of computer vision tasks. However, while this technique is incorporated in existing works, there still lacks a comprehensive investigation on variants of multi-scale convolution in image super resolution. In this work, we present a unified formulation over widely-used multi-scale structures. With this framework, we systematically explore the two factors of multi-scale convolution -- feature propagation and cross-scale communication. Based on the investigation, we propose a generic and efficient multi-scale convolution unit -- Multi-Scale cross-Scale Share-weights convolution (MS$^3$-Conv).
Extensive experiments demonstrate that the proposed MS$^3$-Conv can achieve better SR performance than the standard convolution with less parameters and computational cost. 
Beyond quantitative analysis, we comprehensively study the visual quality, which show that MS$^3$-Conv behave better to recover high-frequency details.
\keywords{Multi-Scale Convolution; Super-Resolution}
\end{abstract}

\section{Introduction}

Image super resolution (SR) is inherently a multi-scale problem, where the output high-resolution (HR) image scale-larger than the input low-resolution (LR) image. Multi-scale technique has also been applied in SR in several ways. Prior to the success of deep-learning-based methods, its first application can be traced back to the self-similarity method, which employs similar patches in an image pyramid downscaled by the LR image. Later in the deep learning era, resizing the training images with multiple scales (e.g., 0.6-0.9) has been a general pre-processing step for data augmentation. Advanced network structures, like LapSRN \cite{lai2017deep} and U-net \cite{ronneberger2015u}, have also incorporated the idea of multi-scale in middle features. While being an inevitable part in SR, multi-scale itself has been few deeply studied in previous literature. In this work, we systematically investigate the design and utilization of multi-scale convolution in SR, and propose a new and efficient multi-scale architecture with faster speed and better performance.

\begin{figure}[t]
  \centering
  \includegraphics[width=0.98\linewidth]{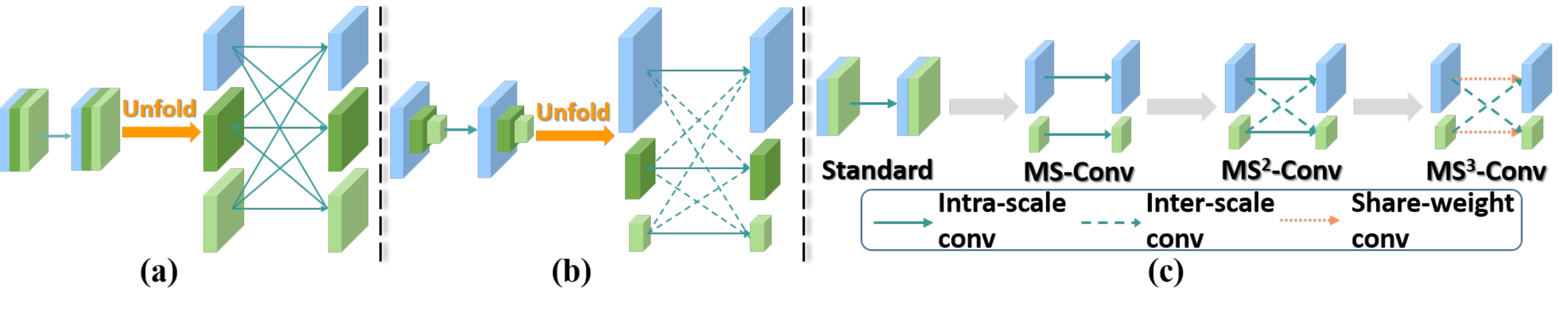}
  \vskip -0.3cm
  \caption{\textbf{(a)} Standard convolution can be unfolded into multiple branches. \textbf{(b)} multi-scale convolution can also be unfolded and grouped by corresponding scales. \textbf{(c)} Evolution of variants of multi-scale convolution.}
  \label{fig:msconv}
  \vskip -0.5cm
\end{figure}

To facilitate the investigation, we propose a novel formulation to cover different variants of multi-scale structures. Specifically, we formulate the widely-used structures -- U-net \cite{ronneberger2015u}, Octave convolution \cite{chen2019drop}, Multi-grid convolution \cite{ke2017multigrid} in a unified framework (see Eqn. \ref{primary eqn}) with different transformations. The formulation also reveals that the performance of multi-scale convolution depends on two factors – feature propagation and cross-scale communication. The first one determines how each scale works, while the second one controls the information flow. We then progressively modify the transformation function and explore the most efficient design. The evolution of modifications is illustrated in Figure \ref{fig:msconv}(c).

\begin{figure}[t]
  \centering
  \includegraphics[width=0.98\linewidth]{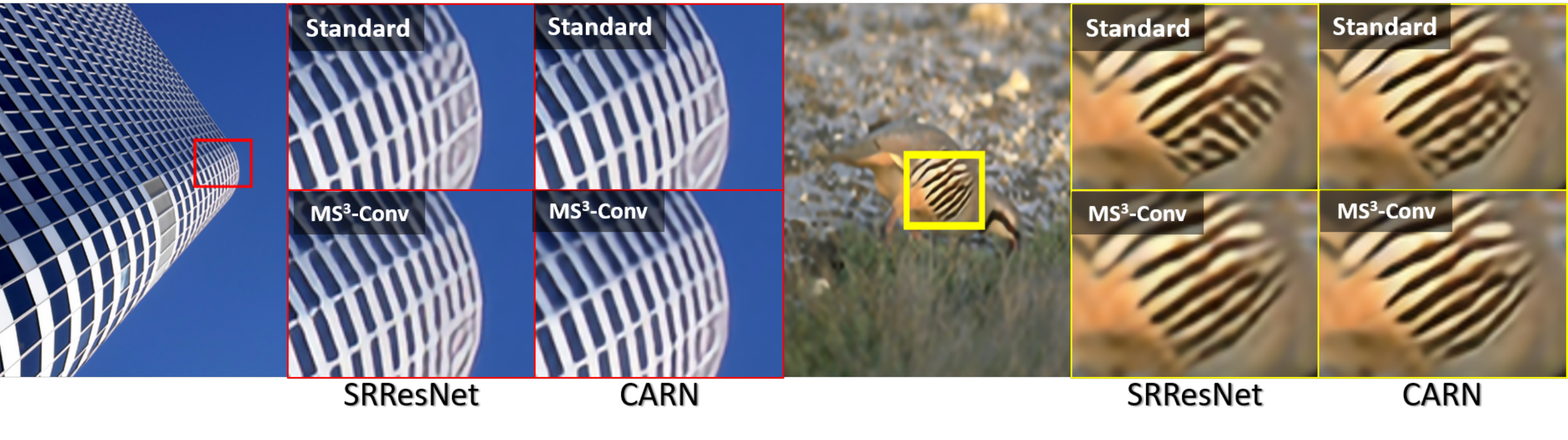}
  \vskip -0.4cm
  \caption{Results of the standard convolution (the top row) and MS$^3$-Conv (the bottom row) on SRResNet \cite{ledig2017photo} and CARN \cite{ahn2018fast} backbone. MS$^3$-Conv on both backbone networks can recover the lattice and the stripe pattern, while the standard convolution generates totally wrong structures.}
  \label{fig:forepic}
  \vskip -0.5cm
\end{figure}

To begin with, the basic multi-scale convolution is to split the original convolution into multi-branches and aggregate these branches at the end of the network. Features will be propagated in high-/low-scales separately. This simplest multi-scale version (MS-Conv), which is widely used in high-level vision tasks \cite{lin2017feature, zhao2017pyramid, huang2017multi}, could undoubtedly reduce the computation complexity, but also bring severe performance degradation in low-level task. Then we add cross-scale communication paths to allow information flow between different branches. We experimentally find that bi-directional cross-scale connections can significantly improve the performance. We name this improved variant as Multi-Scale$+$cross-Scale convolution (MS$^2$-Conv). To further save parameters and computation cost, we adopt share weights strategy on high-/low-scale filters, and use smaller cross-scale filters. The final version, denoted as Multi-Scale$+$cross-Scale$+$Share-weights convolution (MS$^3$-Conv), can achieve similar PSNR as its single-scale baseline but with only $67\%$ of computation complexity and $75\%$ of total parameters. Extensive experiments evaluate the effectiveness of each component. 

Except for quantitative analysis, we also comprehensively test the visual quality between different variants. Different from previous works, we surprisingly find that the outputs of multi-scale and single-scale networks exhibit completely different visual effects, even with similar PSNR values. This is mainly due to their preference to different structures. In general, multi-scale convolutions are superior to recover high-frequency details (e.g., dense grids). For example, in Figure \ref{fig:forepic}, we observe that the MS$^3$-Conv version on both backbone networks could fairly reconstruct the dense lines and grids, while the baseline networks cannot. In some flat regions, we notice that large PSNR gaps do not indicate significant visual differences. With similar quantitative results, we will favor the qualitative performance of multi-scale networks.

In addition to the above advantages, multi-scale convolution is a generic and plug-and-play technique, which can be easily equipped with existing network structures. We have transformed the state-of-the-art networks -- CARN  \cite{ahn2018fast}, SRResNet \cite{ledig2017photo} to their multi-scale versions and compare with the original models on an equal footage. Experiments show that the proposed MS$^3$-Conv can significantly reduce the computation cost and parameters with little sacrifice of performance. The contributions of this work are three-fold:

\begin{enumerate}
\item We propose a unified explanation for understanding and designing multi-scale convolution networks. In this framework, existing multi-scale structures share the same formulation but differ in transformation functions.
\item Based on this unified framework, we conduct a systematic investigation on different variants of multi-scale convolutions. Ablation study demonstrates that the proposed MS$^3$-Conv is more memory and computation efficient than single-scale baseline.
\item We have comprehensively studied the visual quality of various image regions. Experiments show that multi-scale networks are superior to reconstruct high-frequency details.
\end{enumerate}

\section{Related work}
\textbf{Image Super Resolution.}
With the seminal exploration of employing deep learning in SR task \cite{dong2014learning,dong2016image}, the variational approaches with deep neural networks have been dominated single image SR.
Subsequently, VDSR \cite{kim2016accurate} further improves the performance by introducing \textit{global residual learning} to alleviate the training difficulty.
Instead of using predefined upsampling operators as in previous works, FSRCNN \cite{dong2016accelerating} takes the original LR images as input and upscales the feature maps by learning a deconvolution layer (a.k.a. transposed convolution layer) at the very end of the networks.
Similarly, ESPCN \cite{shi2016real} proposes an efficient sub-pixel convolution layer sharing the same motivation as FSRCNN.
The learning upsampling operators are also applied in SRResNet \cite{ledig2017photo}, and EDSR \cite{lim2017enhanced} further expand the network size and dramatically boosts the performance.
Recently, several works, including SRDenseNet \cite{tong2017image},  ResidualDenseNet \cite{zhang2018residual}, RRDB \cite{wang2018esrgan} demonstrate the efficiency of dense connection, where each layer utilizes information from all preceding layers.
CARN \cite{ahn2018fast} constructs a compact and efficient ResNet using cascading mechanism to incorporate multi-layer information.
In addition, RCAN \cite{zhang2018image} explores deeper architecture with channel attention mechanism.
Wang \etal \cite{wang2018recovering} proposes a novel spatial feature transform layer to incorporate the semantic prior and \cite{gu2019blind} proposes IKC method to cope with blind SR problem.
\\
\textbf{Multi-scale Representations.}
Multi-scale representation has exhibited great success in multiple computer vision tasks. FPN \cite{lin2017feature} and PSP \cite{zhao2017pyramid} merge convolutional features from different depths at the end of the networks for object detection and segmentation tasks. MSDNet \cite{huang2017multi} and HR-Nets \cite{sun2019high}, proposed carefully designed network architectures that contain multiple branches where each branch has it own spatial resolution. Hourglass \cite{newell2016stacked} combines low-level features in the high-to-low process into the same resolution features through skip connections. The bL-Net \cite{chen2018big} and Elastic \cite{wang2018elastic} adopt similar idea, but are designed as a replacement of residual block for ResNet \cite{he2016identity} and thus are more flexible and easier to use. Multi-grid CNNs \cite{ke2017multigrid} propose a multi-grid pyramid feature representation and define the MG-Conv operator that can be integrated throughout a network. Oct-Conv \cite{chen2019drop} shares a similar idea with MG-Conv but the motivation is to reduce spatial redundancy.


\section{Multi-Scale Networks}
\subsection{Formulation}
We first present a unified formulation for multi-scale convolution from multi-branch view. To explore the relations of different multi-scale networks, we regard multi-scale feature maps as multiple parallel branches, as shown in Figure \ref{fig:msconv}(b). Throughout this section, we specialize our analysis to a division of spatial resolution by a power of 2. Formally, a vanilla convolution layer can be recast as a combination of \textit{splitting}, \textit{transforming} and \textit{aggregating} operations. From the view of multi-branch networks, we denote $X_H$ and $X_L$ as the \textit{split} input features at higher (finer) and lower (coarser) scale, respectively. For output features of each branch ($Y_H$, $Y_L$), the \textit{transformation} function $f(\cdot)$ can be factorized into four terms that transform the gathered information to the corresponding output branch. Specifically, $f_{HH}$ and $f_{LL}$ denote intra-scale transformations, while $f_{HL}$ and $f_{LH}$ represent inter-scale ones. Then the transformations are \textit{aggregated} by summation, i.e., $Y_H = f_{HH}(X_H) + f_{LH}(X_L)$. The multi-scale convolution can be formulated as follows.
\begin{eqnarray}
\label{primary eqn}
\left[ \begin{array}{c}
Y_H \\ Y_L
\end{array}\right]=
\left[ \begin{array}{cc}
f_{HH} & f_{LH} \\f_{HL} & f_{LL}
\end{array}\right]
\left[ \begin{array}{c}
X_H \\ X_L\end{array}\right]
\end{eqnarray}
Eqn. (\ref{primary eqn}) encapsulates the aggregating transformation performed on the input feature maps.
The multi-scale convolution is depicted in Figure \ref{fig:msconv} (b). The unfold version resembles a multi-branch full-connection network with both inter- and intra-scale transformations. As illustrated in Figure \ref{fig:msconv} (a), the left regular convolution is equivalent to the right multi-branch convolutions, where $X_H$ and $X_L$ are of the same scale. For multi-scale convolution networks, the inter-scale transformations are coupled with up- or down-sample operators to match the spatial resolution of the gathered features. This formula could be easily extended to networks with more than $2$ parallel branches (scales). 

This formulation provides us a principled way to understand multi-scale convolution networks. It suggests that the performance of multi-scale convolution depends on two factors: (i) the intra-scale transformations $f_{HH}, f_{LL}$ that are responsible for feature propagation, and (ii) the inter-scale transformations $f_{HL}, f_{LH}$ that represent cross-scale communications. Next, we analyze several widely-used multi-scale convolution networks in this framework.

\textbf{U-Net} \cite{ronneberger2015u}. We consider the simplest format of U-Net architecture where the feature maps are down-sampled by scale factor $2$. The transformation function is defined as:
\begin{eqnarray}
f=
\left[ \begin{array}{cc}
\mathbf{I} & \mathbf{0} \\ \mathbf{0} & \mathbf{W}_{LL}
\end{array}\right],
\end{eqnarray}
where $\mathbf{I}$ denotes identity mapping. This means that the high-resolution information is conveyed by skip connections and low-resolution branch is transformed by a set of convolution filters, and there is no inter-scale information exchange between branches.

\textbf{Octave Convolution} \cite{chen2019drop}. The transformation matrix of Octave Convolution can be interpreted as:
\begin{eqnarray}
f=
\left[ \begin{array}{cc}
\mathbf{W}_{HH} & \uparrow\circ\mathbf{W}_{LH} \\ \mathbf{W}_{HL}\circ\downarrow_{avg.} & \mathbf{W}_{LL}
\end{array}\right],
\end{eqnarray}
where $\uparrow\circ\mathbf{W}_{LH}$ is a composition function of nearest neighbor upsampler $\uparrow$ and convolution function $\mathbf{W}_{LH}$, and $\downarrow_{avg.}$ represents average pooling with stride 2. Octave Convolution performs both intra- and inter-scale transformations on feature maps at finer and coarser scales to reduce memory and computation cost.

\textbf{Multi-Grid Convolution} \cite{ke2017multigrid}. Multi-Grid Convolution (MG-Conv) is designed for exploiting multi-scale features. For feature maps divided into $3$ scales $X=\{X_H; X_M; X_L\}$, MG-Conv performs the aggregated transforms as
\begin{eqnarray}
\Biggl[ \begin{smallmatrix}
Y_H \\ Y_M \\ Y_L
\end{smallmatrix}\Biggr]=
\left[ \begin{smallmatrix}
\mathbf{W}_{HH} & \mathbf{W}_{MH}\circ \uparrow & \mathbf{0}\\
\mathbf{W}_{HM}\circ\downarrow_{max} & \mathbf{W}_{MM} & \mathbf{W}_{LM}\circ \uparrow\\
\mathbf{0} & \mathbf{W}_{ML}\circ\downarrow_{max} & \mathbf{W}_{LL}
\end{smallmatrix}\right]
\left[ \begin{smallmatrix}
X_H \\ X_M \\ X_L\end{smallmatrix}\right].
\end{eqnarray}
Here $\downarrow_{max}$ is a max pooling that facilitates lateral communication from coarser to finer scale.
It is worth noting that information is transmitted to only the neighbor scales, e.g., $X_H$ would not be transformed and aggregated to $Y_L$, and vice versa.

\subsection{Multi-Scale Share-Weights Convolution}

\begin{figure}[t]
\centering
\includegraphics[width=0.98\linewidth]{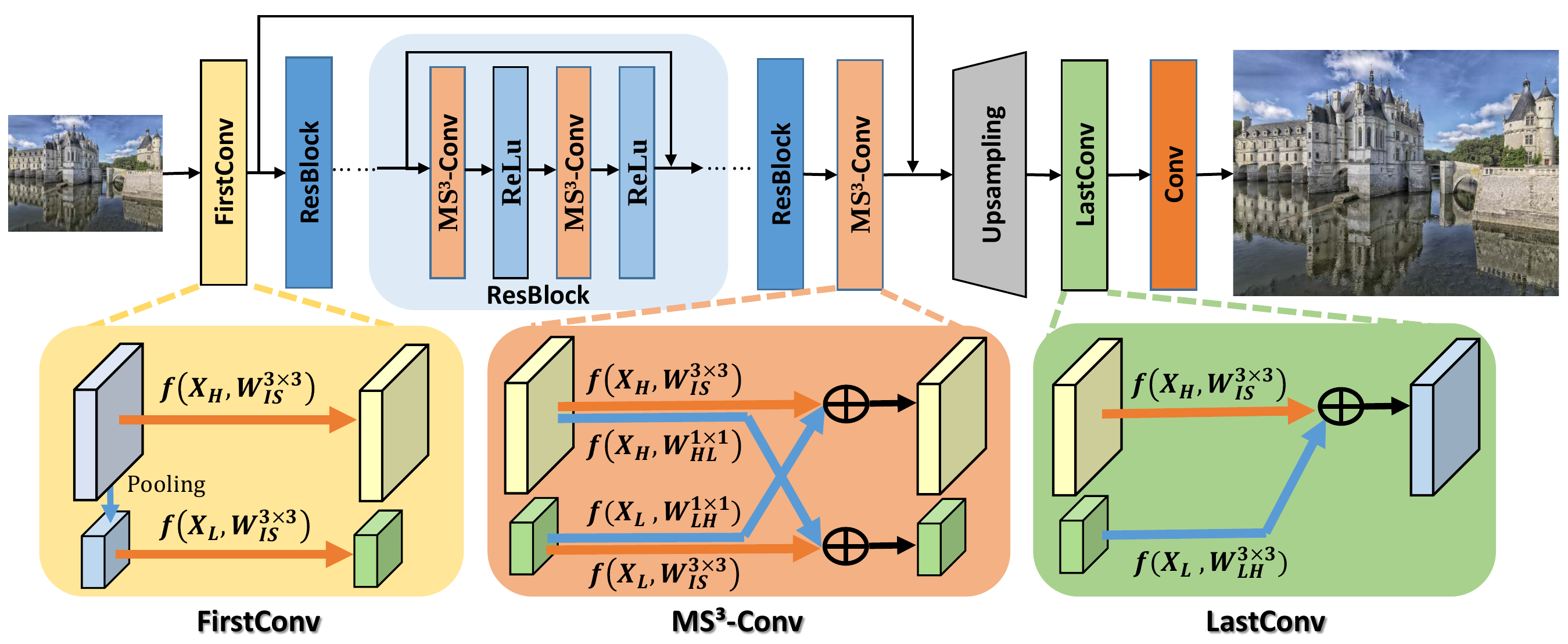}
\vskip -0.2cm
\caption{Illustration of the proposed MS$^3$-Conv for SRResNet. The FirstConv transforms the original-scale image features into two scales. The main network contains two branches and adopts both the inter-scale and intra-scale communications. The LastConv aggregates the two-scale features for final reconstruction. }
\label{fig:resnet}
\vskip -0.4cm
\end{figure}

Our investigation starts from the simplest multi-scale convolution, denoted as MS-Conv, where features will be split into two branches and propagated separately. This modification could reduce the computation complexity, but also cause a severe performance drop. To alleviate this problem, we add cross-scale communication paths to allow information exchange between different branches, which is conceptually similar to OctConv \cite{chen2019drop}. We name this improved variant as Multi-Scale$+$cross-Scale convolution (MS$^2$-Conv).

Based on the above analysis, we further propose a more efficient multi-scale variant, referred to as Multi-Scale cross-Scale Share-weights convolution (MS$^3$-Conv). The key idea is keep the inter-scale communication paths --$ f_{HL}, f_{LH}\neq0$ and adopt share weights strategy for intra-scale transformations -- $ f_{HH}=f_{LL}$. To keep cross-scale communication could lead to significant performance improvement, which has been demonstrated by experiments (see Section \ref{ms-to-ms3}). To share weights is inspired by TridentNet \cite{li2019scale}, which constructs a parallel multi-branch architecture and uses the same transformation parameters for different scales. In addition, we apply $1\times1$ convolutions for inter-scale communications, instead of $3\times3$ ones, to reduce redundant parameters. To sum up, the MS$^3$-Conv performs the transformation as
\begin{eqnarray}
f=
\left[ \begin{array}{cc}
\mathbf{W}_{IS} & \uparrow\circ\mathbf{W}_{LH} \\ \mathbf{W}_{HL}\circ\downarrow_{avg.} & \mathbf{W}_{IS}
\end{array}\right],
\end{eqnarray}
where $\mathbf{W}_{IS}$ represents $3\times3$ convolutions for feature propagation, and $\mathbf{W}_{LH}$ and $\mathbf{W}_{HL}$ are $1\times1$ convolutions for cross-scale communication. Figure \ref{fig:msconv}(c) summarizes the evolution from MS-Conv to MS$^3$-Conv. Figure \ref{fig:resnet} exemplifies how to equip MS$^3$-Conv to SRResNet.

The key differences between MS$^3$-Conv and OctConv \cite{chen2019drop} are: 1) For intra-scale transformation, MS$^3$-Conv uses shared parameters for each scale. The splitting ratio for high-/low-scales is fixed at $0.5$. 2) For inter-scale communication, MS$^3$-Conv adopts two $1\times1$ convolutions, while OctConv uses $3\times3$ convolution.

Compared with TridentNet \cite{li2019scale}, MS$^3$-Conv is formulated as a generic and plug-and-play convolution unit that can be used in most network architectures, while TridentNet constructs multi-branch blocks and incorporates them into the backbone networks. In addition, MS$^3$-Conv adopts convolution for inter-scale communications, while TridentNet contains no communication paths and uses NMS to combine the outputs of different branches in the final stage.

\subsection{Why multi-scale and share-weights work?}
\begin{figure}[t]
\centering
\includegraphics[width=0.98\linewidth]{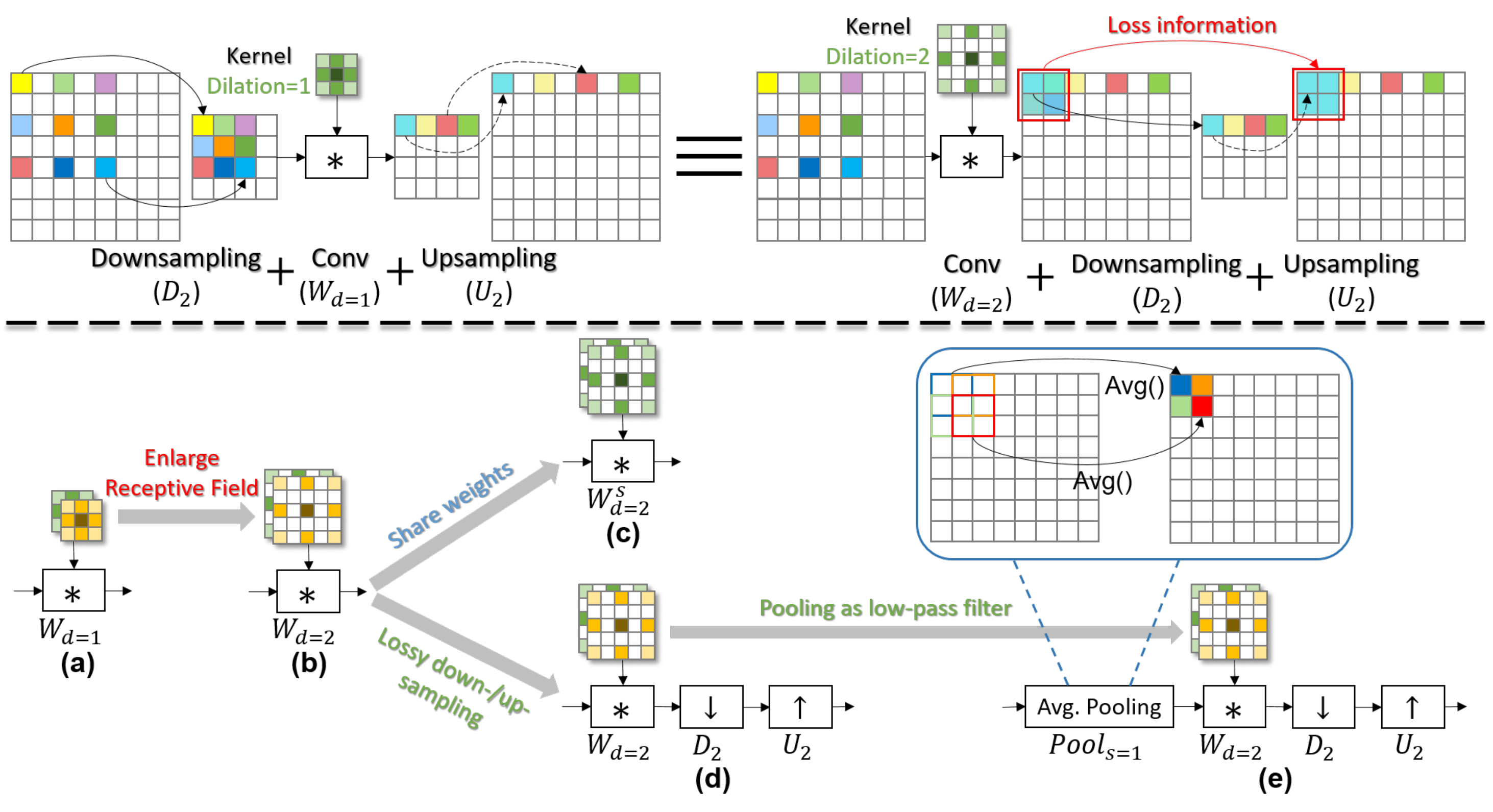}
\caption{\textbf{(Top)} A convolution performed on downsampled feature maps is functionally equivalent to one with dilation rate $2$, followed by lossy function (down-/up-sampling). This equality enables an investigation on multi-scale convolutions in isolation from computation cost.
\textbf{(Bottom)} Relations of various multi-scale convolutions. All cases have the same computational complexity but capture different information with four related factors, including receptive field, share weights, lossy down-/up-sampling, and low-pass filter. See Table \ref{pilot} for PNSR results.}
\label{fig:whywork}
\vskip -0.3cm
\end{figure}

\begin{table}[t]
  \centering
  \caption{Pilot experiments on BSD100 dataset. Several cases are illustrated in Figure \ref{fig:whywork}. This demonstrates that simply increasing receptive field size and the use of lossy down-sampler may not be essential for SR.}
  \begin{tabular}{lccl}
  	\hline
  	Function & Figure & PSNR & Remarks \\\hline
  	$W_{d=1}$ & Fig. \ref{fig:whywork}(a) & 27.58 & -\\
  	$W_{d=2}$ & Fig. \ref{fig:whywork}(b) & 27.57 & Enlarge receptive field\\
  	$W_{d=2}^S$ & Fig. \ref{fig:whywork}(c) & 27.56 & Share weights\\
  	$W_{d=2} - D_2 - U_2$ & Fig. \ref{fig:whywork}(d) & 27.56 & Lossy function\\
  	$Pool_{s=1}-W_{d=2} - D_2 - U_2$ & Fig. \ref{fig:whywork}(e) & 27.57 & Low-pass filter; Compensation\\
  	\hline
  \end{tabular}
  \label{pilot}
  \vskip -0.4cm
\end{table}

Multi-scale representation has long been applied to capture the spatial long-range dependency and spatial redundancy. To study the effects of multi-scale in isolation from spatial redundancy, we bridge multi-scale convolution with dilated convolution \cite{yu2015multi}, which enlarges the receptive field by performing convolution at sparsely sampled locations. With dilated convolutions, different branches of the networks could have the same structure yet have different receptive fields.

We consider the simplest multi-scale representation function $D_2 - W_{d=1} - U_2$, where $D_2$ and $U_2$ are nearest neighbor downsampler and upsampler with scaling factor 2, and $W_{d=1}$ is a convolution filter with dilation rate $1$. As shown in Figure \ref{fig:whywork}, this function can be re-arranged as $W_{d=2} - D_2 - U_2$.
Note that $D_2 - U_2$ is an intrinsically lossy function, and this function does not change the spatial resolution of feature maps. In this view, this multi-scale representation function is a combination of $W_{d=2}$ (Enlarge receptive field) and $D_2 - U_2$ (lossy down-/up-sampling). We also study a case that two branches share the same parameters but with different dilation rates, denoted as $W_{d=2}^S$. 

On top of the simplest case, let us consider a more popular multi-scale structure $Pool_{s=2} - W_{d=1} - U_2$, where $Pool_{s=2}$ denotes average pooling with stride $2$. Similar to the above formula, this function can be recast as $Pool_{s=1} - W_{d=2} - D_2 - U_2$. Here average pooling before convolution filter serve as a low-pass filter and compensate information loss derived from down-/up-sampling part. 

We conduct our pilot experiment for the following five cases in Table \ref{pilot}. The results demonstrate that receptive field size and lossy down-sampler may not be essential for SR. We conjecture this may be partly due to spatial redundancy of images. Moreover, compared to transformations that share weights at each branch, more representation power (parameters) only leads to negligible improvement. Since it would not cause a performance drop, it is reasonable to distribute convolution filters to low-resolution feature maps to reduce computation complexity and use share-weights branches at different scales to reduce memory overhead.

\subsection{Analysis of Visual Quality}
We also comprehensively analyze visual quality between different variants of convolution. In general, multi-scale convolutions are superior to recover high-frequency details (e.g., dense grids). For example, in Figure \ref{fig:msconv}, we observe that the MS$^3$-Conv version on both backbone networks could fairly reconstruct the dense lines and grids, while the baseline networks cannot. In Figure \ref{fig:variants}, the standard convolution recover the wrong structure of grids, while all multi-scale convolution variants are able to alleviate this issue and recover the correct structure. Besides comparing high-frequency regions, we find that in some cases MS$^3$-Conv behaves differently in flat area, even with similar perceptual quality (See Figure \ref{fig:comp_smooth}). Indeed, in some flat regions, we notice that their PSNR gaps are large, whereas the visual differences are perceptually insignificant. This could be partly attributed to their preference to different structures. More details are shown in Section \ref{sec:vq}.

\section{Experiments}
\label{exp}
In this section, we present a systematic investigation of various types of multi-scale convolution. We construct a baseline network -- SRResNet \cite{ledig2017photo}, which contains 16 residual blocks and no BatchNorm layers. All convolution layers have $64$ filters. To measure the computation complexity, we adopt the widely-used metric -- FLOPs to calculate the number of multiply-adds. We first describe the implementation details and training settings in Section \ref{settings}. Then we evaluate the effectiveness from MS-Conv to MS$^3$-Conv and the choice of the number of branches in Section \ref{ms-to-ms3}. Finally, Section \ref{cnns} compares the results of MS$^3$-Conv with standard convolution on other state-of-the-art backbone networks.

\subsection{Implementation details}
\label{settings}
Our models are performed with a scaling factor of $\times4$ based on Pytorch \cite{2017-Paszke-p-} framework and are trained with single NVIDIA Titan Xp GPU. All models are trained/tested based on the same implementation as follows.

Following \cite{lim2017enhanced}, we use 800 training image pairs from DIV2K dataset \cite{Agustsson_2017_CVPR_Workshops} as training set.
To fully utilize the dataset, we also perform data augmentation. To prepare training data, we first crop the HR images into a set of $480\times480$ sub-images with a stride $240$. 
During training, a patch of size $128\times128$ as ground-truth data is randomly cropped from a $480\times480$ sub-image and subsequently down-sampled with bicubic kernel as the LR image. In addition, the training data is augmented with random horizontal/vertical flips and 90 rotations. 
For testing, the evaluation is conducted and compared on standard benchmark datasets: Set5 \cite{bevilacqua2012low}, Set14 \cite{yang2010image}, B100 \cite{martin2001database}, Urban100 \cite{huang2015single}, DIV2K \cite{Agustsson_2017_CVPR_Workshops}, with PSNR criteria on Y channel (i.e., luminance) of transformed YCbCr space. The mini-batch size is $16$ and we train our model with ADAM \cite{kingma2014adam} optimizer by setting $\beta_1=0.9$, $\beta_2=0.999$, and $\epsilon=10^{-8}$. The learning rate is initialized as $2\times10^{-4}$ and then decayed by half every $2.5\time 10^5$ iterations for totally $1\times 10^6$ iterations. We use $\ell_1$ loss instead of $\ell_2$ as suggested in \cite{lim2017enhanced}.

\subsection{Ablation study}
\label{ms-to-ms3}
\textbf{Effectiveness of Multi-Scale (MS-Conv).}
We first analyze how the multi-scale convolution affects the performance in comparison with standard convolution. To integrate multi-scale convolution into the baseline network, we replace a standard convolution layer with its MS-Conv counterpart (see Figure \ref{fig:msconv}(c)), while the main topology of networks and other configurations remain unchanged for a fair comparison. For each branch (scale), we adopt a $32$-channel $3\times3$ convolution operator. 
Their PSNR values on DIV2K test set are shown in Table \ref{connections}. With half of the feature maps compressed to the lower scale, the computation complexity drops from $42.75$G to $16.70$G. Such a heavy compression also results in a clear PSNR drop -- $0.16$ dB. This shows that MS-Conv can realize a trade-off between performance and complexity. Note that the sacrifice of performance is acceptable in some real applications that prefer fast speed.

\textbf{Effectiveness of Cross-Scale Communications (MS$^2$-Conv).}
\begin{table}[t]
  \centering
  \caption{Comparison of the standard and multi-scale convolution on DIV2K. ``LH'' and ``HL'' are denoted as low-to-high and high-to-low communication path, respectively.}
  \begin{tabular}{ccccccc}
  	\hline
  	\multirow{2}{*}{Conv} & \multirow{2}{*}{Standard} & \multirow{2}{*}{MS-Conv} & MS$^2$-Conv & MS$^2$-Conv & \multirow{2}{*}{MS$^2$-Conv} \\
  	&&&w/o LH &w/o HL &\\\hline
  	Transformation & - & 
  	$\Bigl[ \begin{smallmatrix} \mathbf{W}_{HH} & \mathbf{0} \\\mathbf{0} & \mathbf{W}_{LL} \end{smallmatrix} \Bigr]$ &
  	$\Bigl[ \begin{smallmatrix} \mathbf{W}_{HH} & \mathbf{0} \\ f_{HL} & \mathbf{W}_{LL} \end{smallmatrix} \Bigr]$ & 
  	$\Bigl[ \begin{smallmatrix} \mathbf{W}_{HH} & f_{LH} \\\mathbf{0} & \mathbf{W}_{LL} \end{smallmatrix} \Bigr]$ &
  	$\Bigl[ \begin{smallmatrix} \mathbf{W}_{HH} & f_{LH} \\ f_{HL} & \mathbf{W}_{LL} \end{smallmatrix} \Bigr]$ \\
  	FLOPs (G) & 42.76 & 16.70 & 18.74 & 18.74 & 20.78 \\
  	Params. (M) & 1.59 & 0.82 & 1.21 & 1.21 & 1.59\\
  	PSNR & 30.47 & 30.31 & 30.33 & 30.29 & 30.38 \\
  	\hline
  \end{tabular}
  \label{connections}
  \vskip -0.3cm
\end{table}
To further improve the performance, we evaluate the effects of cross-scale communications. For the transformation function, we set $f_{LH}$ and $f_{HL}$ to be $\uparrow\circ\mathbf{W}_{LH}$ and $\mathbf{W}_{HL}\circ\downarrow_{avg.}$, respectively, which is suggest in OctConv \cite{chen2019drop}. Similar to the feature propagation convolution, $\mathbf{W}_{LH}$ and $\mathbf{W}_{HL}$ are $3\times3$ convolution layers with $32$ filters. We also test the effectiveness of uni-directional cross-scale communication by removing either path.

The comparisons of variants are summarized in Table \ref{connections}. For example, there is no high-to-low path in MS$^2$-Conv w/o HL, thus the transformation $f_{HL}$ is $\mathbf{0}$ in the transformation matrix. The results suggest the importance of both inter-scale communication paths (high-to-low and low-to-high), as removing either of them leads to a performance drop. In particular, comparing $2$rd, $3$th and $4$th column in Table \ref{connections}, we find that networks with single communication path and no path obtain similar performance (the difference is less than $0.02$ dB), which indicates that bi-directional cross-scale connections are essential for information flow. In addition, MS$^2$-Conv improves $0.07$ dB over MS-Conv with around 4 GFLOPs increase.

\begin{figure}[t]
\centering
\begin{minipage}[b]{0.48\textwidth}
\centering
\includegraphics[width=0.9\linewidth]{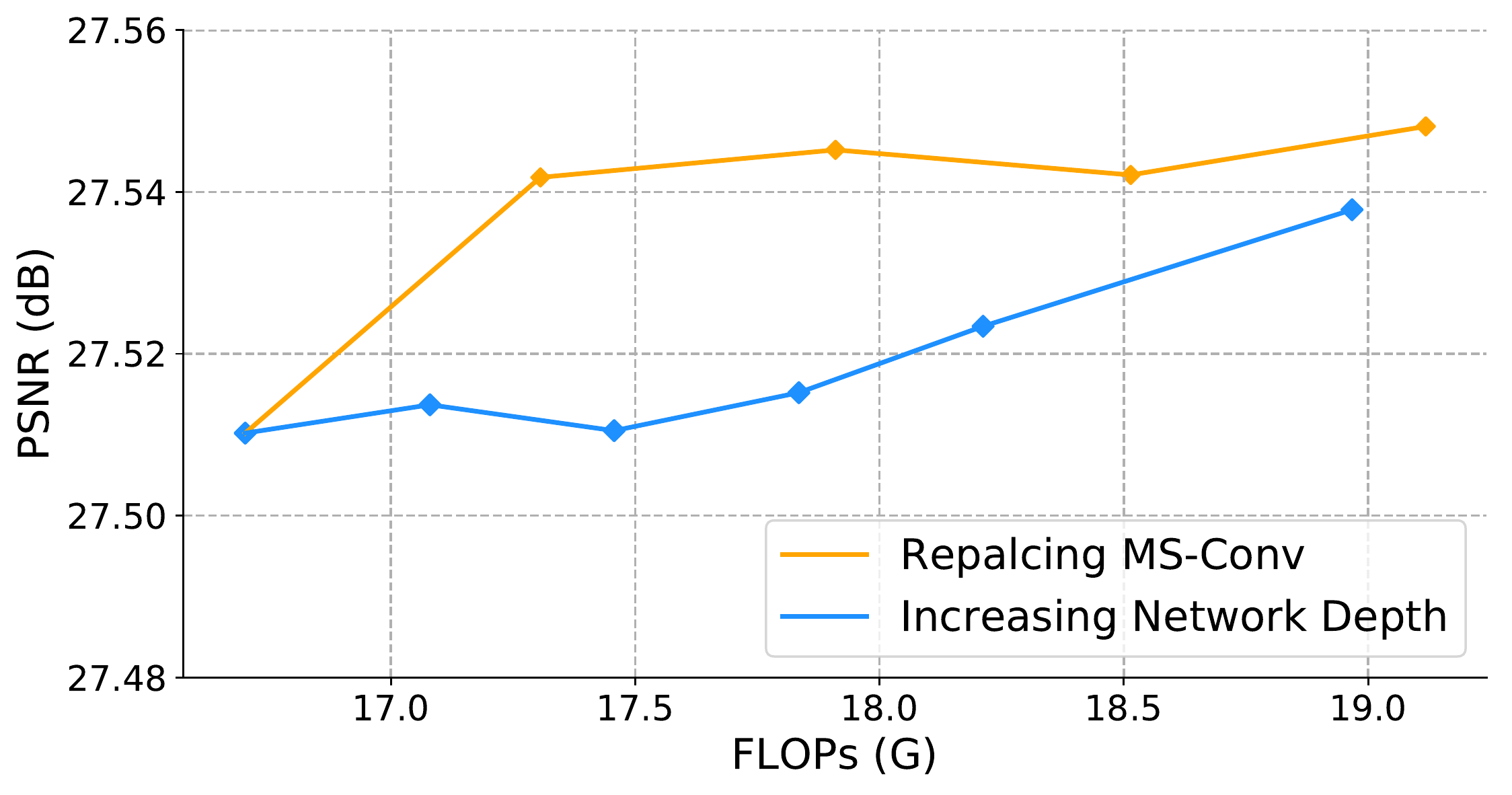}
\caption{PSNR \textit{vs.} FLOPs on BSD100. Two cases of increasing complexity.}
\label{fig:depth-connections}
\end{minipage}
\begin{minipage}[b]{0.48\textwidth}
\centering
\captionsetup{type=table} 
  \begin{tabular}{cccc}
  	\hline
  	Branches & FLOPs (G) & Params (M) & PSNR\\\hline
  	1 & 25.85 & 0.56 & 30.17\\
  	2 & 17.16 & 0.51 & 30.30\\
  	3 & 18.19 & 0.71 & \textbf{30.32}\\
  	4 & 18.48 & 1.00 & 30.29\\
  	\hline
  \end{tabular}
\caption{Comparison of networks with different numbers of scales on DIV2K. The best result is \textbf{highlighted}.}\label{num-branch}
\end{minipage}
\vskip -0.4cm
\end{figure}

To study whether the performance improvement comes from the increased complexity or cross-scale communication, we conduct a series of experiments that start from the MS-Conv network (see Figure \ref{fig:msconv}(c)). We compare two cases of increasing complexity: In the first case, we gradually increase network depth. In the second case, we fix the network depth and gradually replace MS-Conv with MS$^2$-Conv (see Figure \ref{fig:msconv}(c)). We train and evaluate a series of networks under these changes. Figure \ref{fig:depth-connections} shows the comparisons of these two cases on PSNR and FLOPs. Though the difference is relatively minor, we can observe a stable trend that adding cross-scale communication is more effective than increasing depth. These results suggest that cross-scale communication paths facilitate information propagation through networks.

\begin{figure}[t]
\centering
\begin{subfigure}[b]{0.48\linewidth}
\centering
\includegraphics[width=0.95\linewidth]{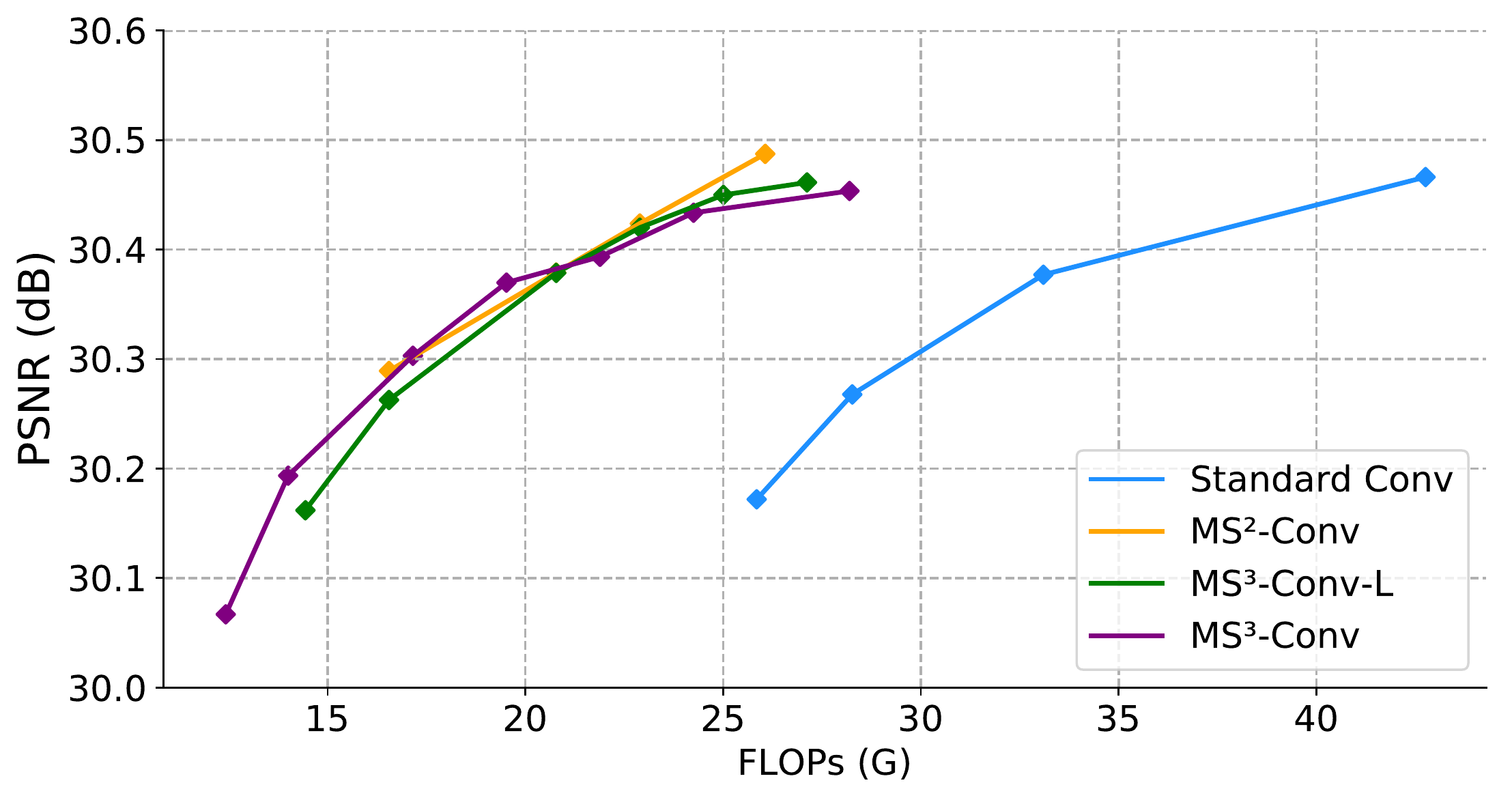}
\vskip -0.2cm
\caption{PSNR \textit{vs.} FLOPs.}
\end{subfigure}
\begin{subfigure}[b]{0.48\linewidth}
\centering
\includegraphics[width=0.95\linewidth]{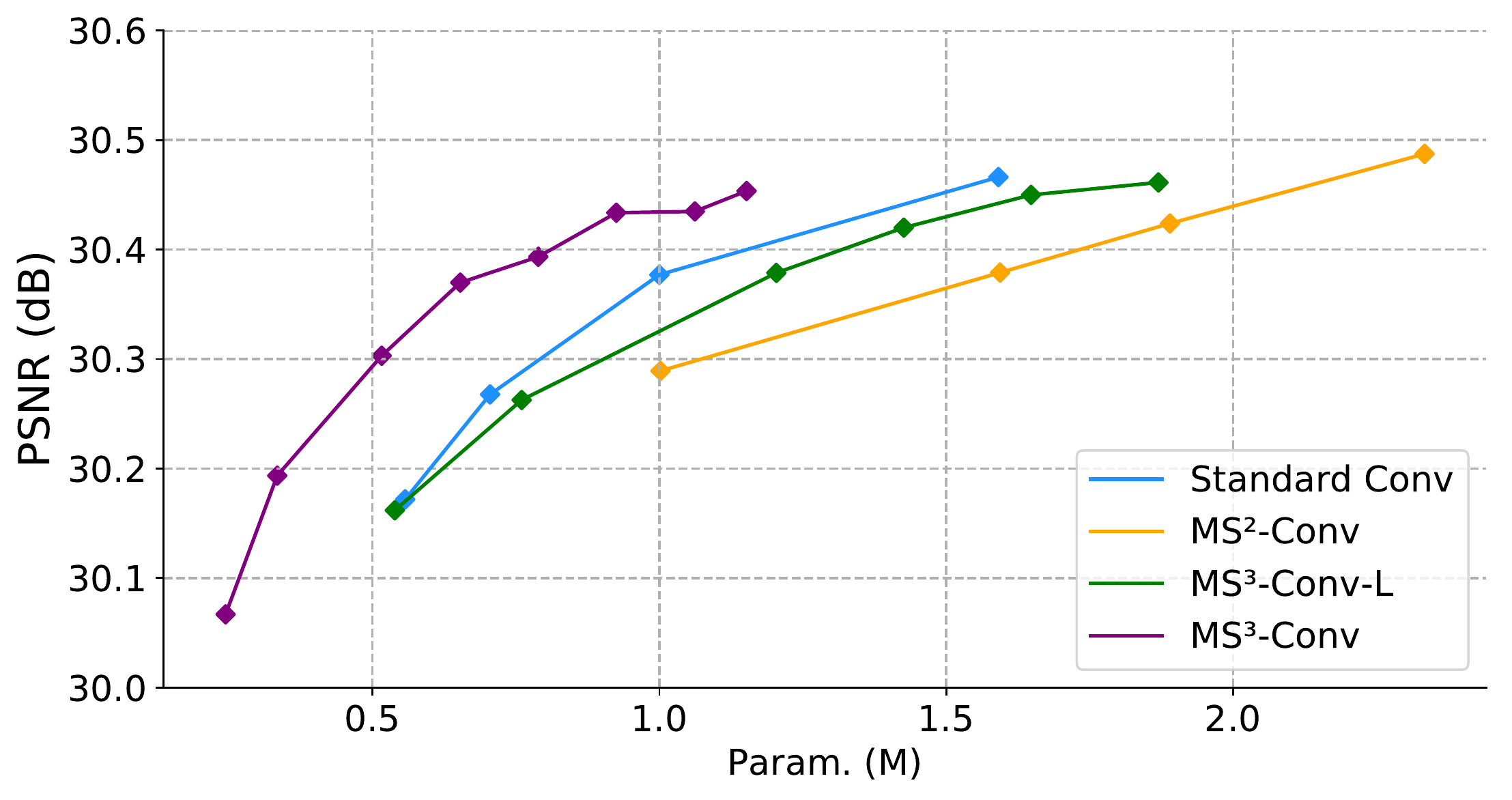}
\vskip -0.2cm
\caption{PSNR \textit{vs.} Param.}
\end{subfigure}
\vskip -0.3cm
\caption{Illustrating the efficiencies of the standard convolution and multi-scale convolution variants. \textbf{(a)} All multi-scale networks achieve the similar performance to standard conv networks with far less computation cost. \textbf{(b)} MS$^3$-Conv demonstrates the best memory-efficiency.}
\label{fig:ms2-conv}
\vskip -0.4cm
\end{figure}

\textbf{Effectiveness of Share-Weights (MS$^3$-Conv).}
To further reduce the number of parameters, we adopt share-weights strategy on filters of different scales, and propose the improved version -- MS$^3$-Conv. To compare MS$^3$-Conv with other multi-scale convolution variants, we evaluate the efficiency of a series of networks at different depths. 
Note that the filter size of inter-scale communication in MS$^3$-Conv is $1\times1$. To evaluate whether a larger filter size performs better, we test another series of networks inter-scale filter size $3\times3$, denoted as MS$^3$-Conv-L. We also include baseline networks in the comparison to show the merits of multi-scale networks. Figure \ref{fig:ms2-conv} shows how the performance of different variants changes when networks go deeper. Comparisons of both PSNR vs. Params and PSNR vs. FLOPs are included.

In Figure \ref{fig:ms2-conv}(a), MS$^2$-Conv, MS$^3$-Conv and MS$^3$-Conv-L exhibit similar trend when going deeper and all multi-scale networks perform consistently better than the baseline networks. Particularly, multi-scale networks can achieve nearly identical performance to the baseline network with only $61\%$ computation complexity,  which demonstrates that multi-scale convolutions can fully utilize the representation power. Moreover, with around $26$ GFLOPs, all multi-scale convolution networks improve over the standard networks by up to $0.3$ dB, which indicates that multi-scale convolutions are more computation efficient.

In Figure \ref{fig:ms2-conv}(b), it can be observed that MS$^3$-Conv is more memory-efficient than MS$^2$-Conv and MS$^3$-Conv-L, as the purple curve remains above the green and yellow ones. To achieve similar peak performance with the baseline network, MS$^3$-Conv,  MS$^3$-Conv-L and MS$^2$-Conv require 1.15M, 1.87M, and 2.33M parameters, respectively. This indicates that enlarging the filter size of inter-scale communication paths only leads to a minor increase, while significantly increasing the model size. Note that both MS$^3$-Conv ($3\times3$) and MS$^2$-Conv require more parameters than the baseline. It also echoes with our motivation to reduce memory costs.

\textbf{Multi-Branch Networks.}
We also conduct more experiments to explore how many parallel branches are required for MS$^3$-Conv. Similar to HRNet \cite{sun2019high}, we conduct multi-scale communications by exchanging the feature information across all parallel branches. Note that the complexities of models with more branches are slightly more than two-branch MS$^3$-Conv due to the extra communication paths.

The results in Table \ref{num-branch} demonstrate that MS$^3$-Conv networks consistently outperform the single-branch method (baseline) with around $0.12$ dB increase. As can be noticed, when adding extra parameters to the networks, three and four branches do not bring significant improvement (less than 0.02 dB) over two branches. Therefore, to achieve better trade-offs between complexity and performance, we choose two branches as our default setting.


\subsection{Results on SOTAs}
\label{cnns}
In this subsection, we equip MS$^3$-Conv on state-of-the-art SR networks -- CARN and SRResNet to generate the corresponding multi-scale version. For fair comparison, we reproduce these networks with the same training settings (see Section \ref{settings}). Table \ref{SOTA} summarizes their results on several SR test sets. In particular, with SRResNet backbone, replacing standard convolution with MS$^3$-Conv could reduce the parameters and computation cost by $67\%$ and $40\%$, respectively. But this setting also results in significant degradation over the backbone network. To compensate the drop, we train a deeper network, denoted as MS$^3$-Conv+, to achieve a similar performance with only two-third computation complexity and three-quarter parameters. Similarly, for CARN backbone networks, it can be found that MS$^3$-Conv+ brings slight improvements ($0.01/0.02$ dB on Urban100/DIV2K test set) over standard convolution, while saving $34\%$ computation cost. Interestingly, we observe that our MS$^3$-Conv+ version SRResNet slightly outperforms CARN backbone on most datasets (e.g., 30.45 \textit{vs.} 30.42 dB on DIV2K), while reducing $20\%$ computation cost. This implies that multi-scale convolution could improve the performance without changing the topology of networks. Comparisons with other state-of-the-art methods can be found in the supplementary material.

\section{Visual Quality Analysis}
\label{sec:vq}

\begin{table}[t]
  \centering
  \caption{Quantitative results on various CNN structures. ``MS$^3$-Conv+'' indicates deeper networks that achieve similar performance with baseline networks. \red{RED}/\blue{BLUE} text represents best/second best results.}
  \begin{tabular}{cccccccccc}
  	\hline
  	Backbone & Conv & FLOPs & Params & Set5 & Set14 & BSD100 & Urban100 & DIV2K\\
	\hline
  	\multirow{3}{*}{SRResNet} & - &$42.76$G &$1.59$M &$\red{32.14}$ &$\red{28.62}$ & $\red{27.58}$ &$\red{26.13}$ & $\red{30.47}$ \\
  	& MS$^3$-Conv & $17.16$G & $0.52$M & $32.01$ & $28.48$ & $27.51$ & $25.84$ & $30.30$ \\
  	& MS$^3$-Conv+ & $28.98$G & $1.19$M & $\blue{32.07}$ & $\blue{28.60}$ & $\blue{27.57}$ & $\blue{26.12}$ & $\blue{30.46}$ \\
  	\hline
  	\multirow{3}{*}{CARN} & - & $35.51$G & $1.15$M & $\red{32.15}$ & $\red{28.58}$ & $\blue{27.57}$ & $\blue{26.03}$ & $\blue{30.42}$ \\
  	& MS$^3$-Conv & $15.06$G & $0.45$M & $32.02$ & $28.50$ & $27.51$ & $25.81$ & $30.30$ \\
  	& MS$^3$-Conv+ & $23.32$G & $1.03$M & $\blue{32.14}$ & $\blue{28.57}$ & $\red{27.58}$ & $\red{26.04}$ & $\red{30.44}$ \\\hline
  \end{tabular}
  \label{SOTA}
  \vskip -0.3cm
\end{table}

\begin{figure}[t]
\centering
\includegraphics[width=0.98\linewidth]{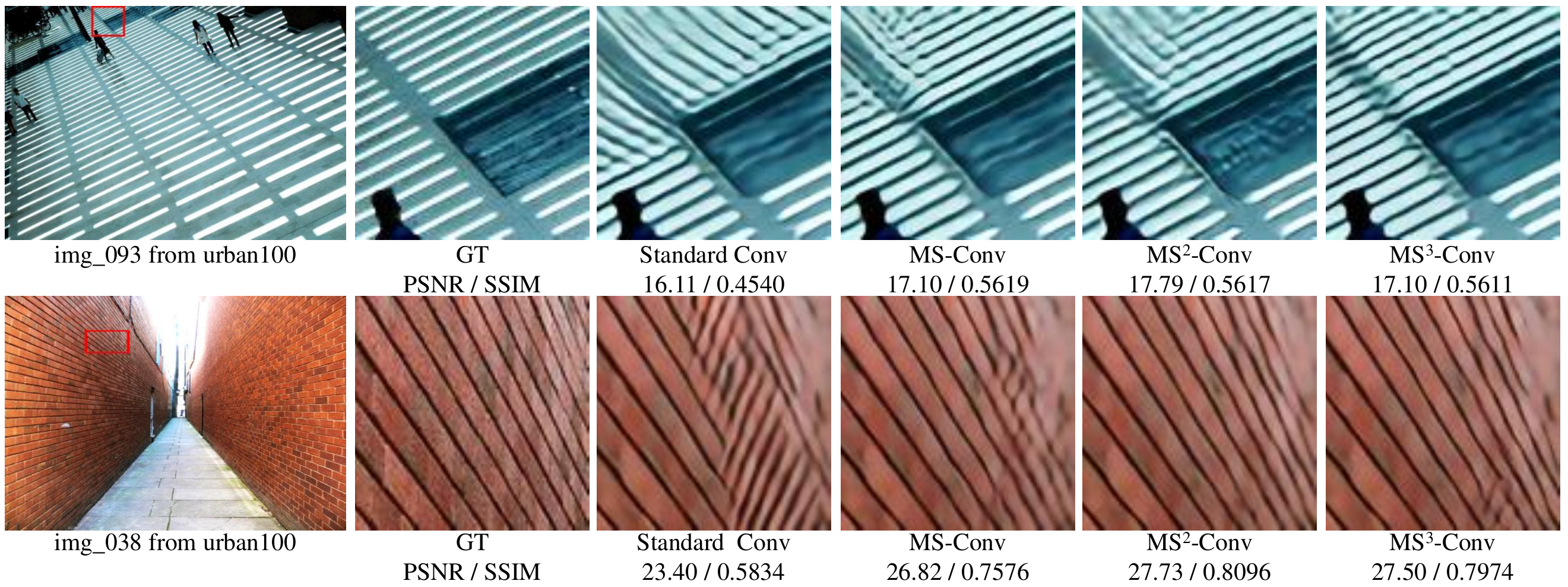}
\vspace{-0.2cm}
\caption{Qualitative comparisons of standard convolution and variants of multi-scale convolution on SRResNet backbone.}
\label{fig:variants}
\vskip -0.4cm
\end{figure}

We compare several variants of multi-scale convolution on public benchmark datasets and present some representative qualitative results in Figure \ref{fig:variants}. PSNR and SSIM are also provided for reference. For image ``img\_093'', the standard convolution recover the wrong structure of zebra-stripes, while all multi-scale convolution variants are able to recover more details and tend to generate the correct structure.
For images ``img\_038'', the standard convolution suffers from ringing artifacts, and MS-Conv, MS$^2$-Conv and MS$^3$-Conv could lead to better visual quality in this region.

\begin{figure}[p]
\centering
\includegraphics[width=0.98\linewidth]{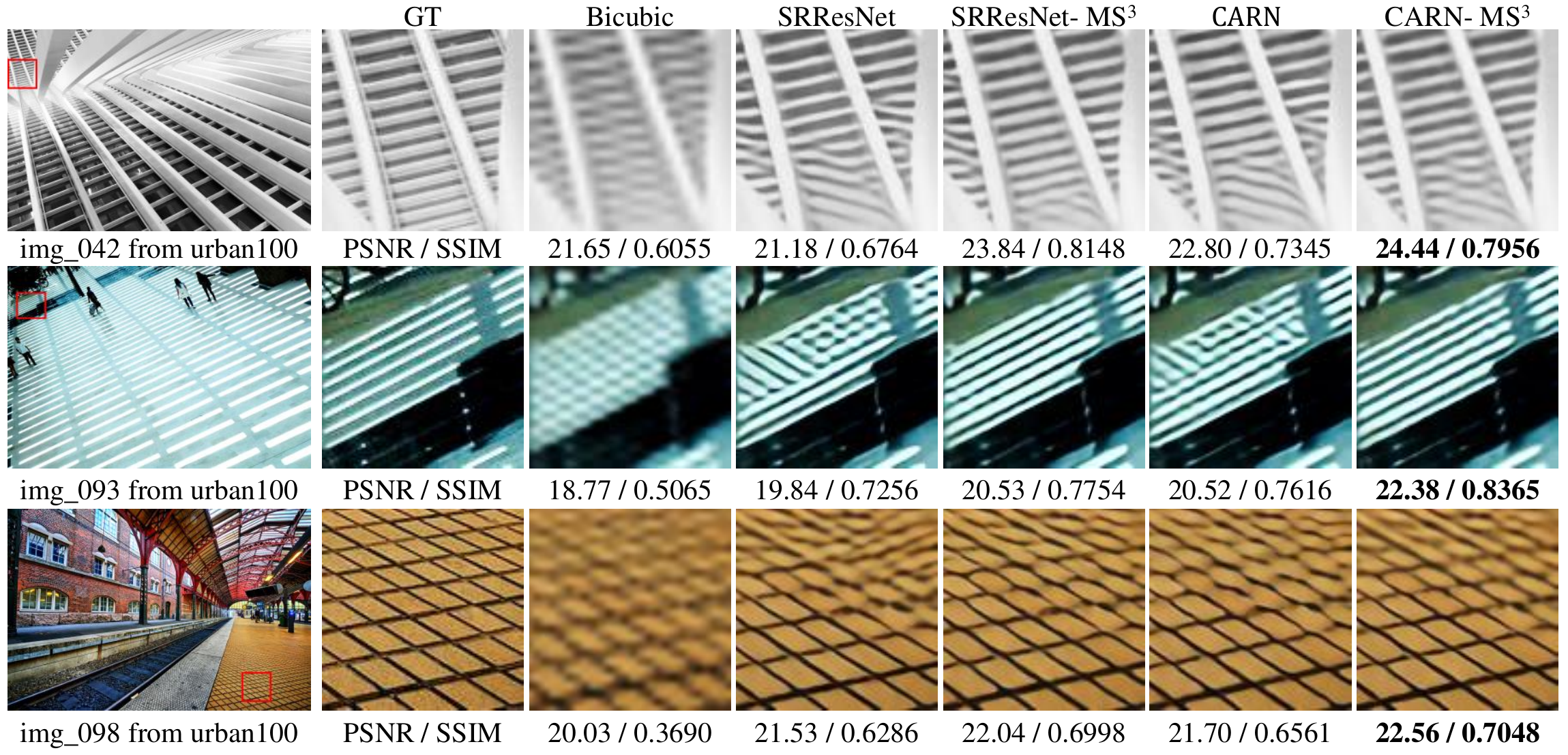}
\caption{Qualitative comparisons on dense-grid region. The best results are \textbf{highlighted}. Multi-scale convolutions are superior to recover high-frequency details, including strips, dense grids, and lattices.}
\label{fig:comp_dense}

\centering
\includegraphics[width=0.98\linewidth]{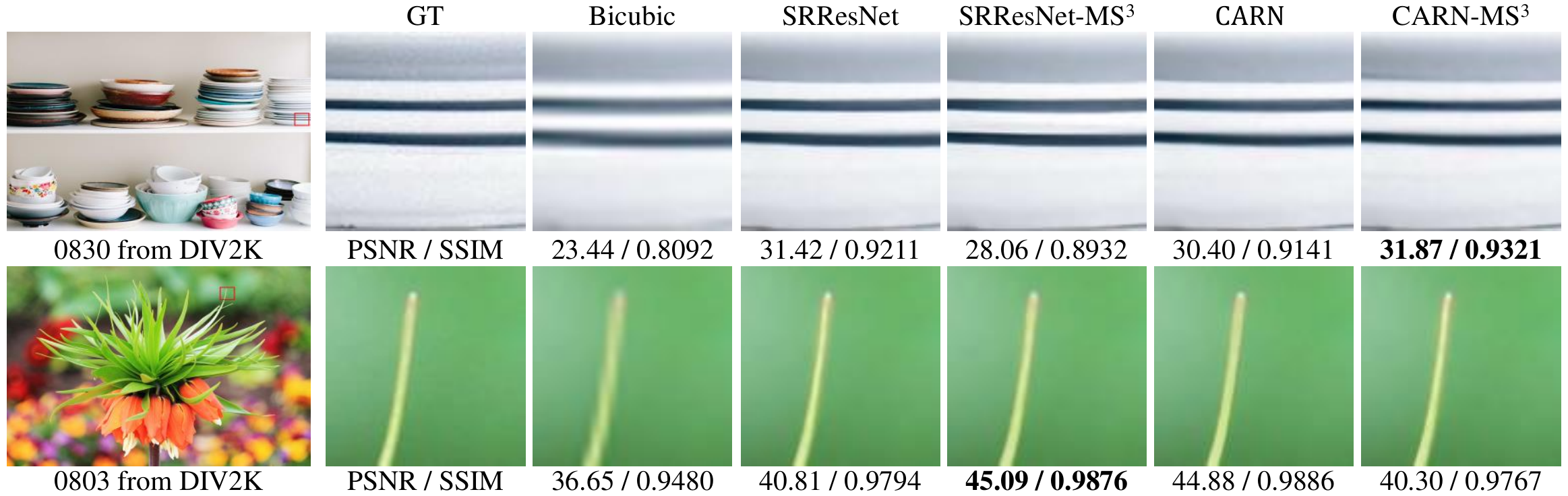}
\caption{Qualitative comparisons on flat region. The best results are \textbf{highlighted}. There exists large PSNR gaps between MS$^3$-Conv and the baseline network on flat region, but the two images are perceptually indistinguishable.}
\label{fig:comp_smooth}

\centering
\includegraphics[width=0.95\linewidth]{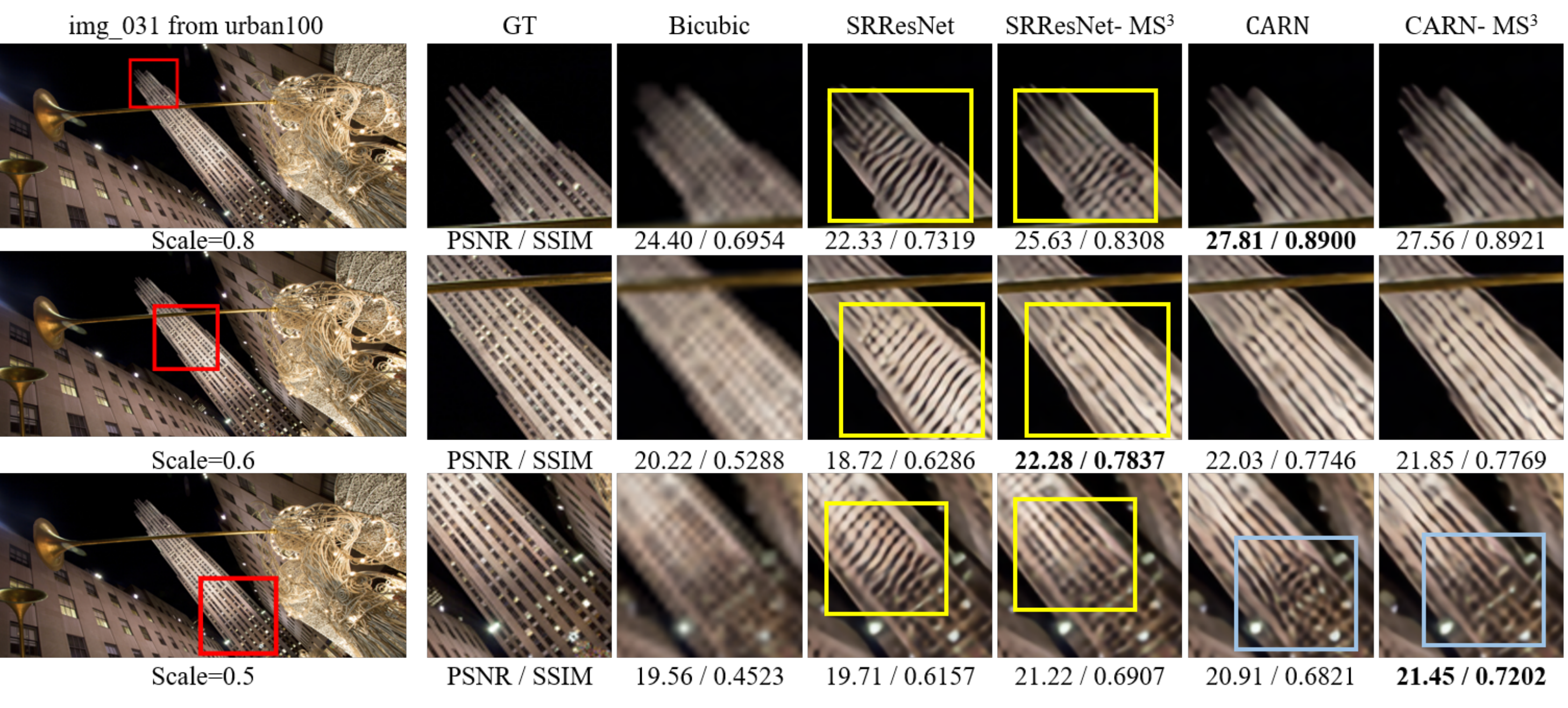}
\caption{Qualitative comparisons on rescaling images, with scaling factor {$0.8$, $0.6$, $0.5$}. All images are rescaled to the same resolution for better visualization. \textbf{Regions of interest in yellow and blue boxes.}}
\label{fig:ms-dataset}
\end{figure}

Beside comparing several variants of SRResNet backbone, we also analyze the visual quality of MS$^3$-Conv with different backbone networks. Figure \ref{fig:comp_dense} shows that multi-scale convolutions are superior to recover high-frequency details (e.g., dense grids). Specifically, both backbone networks fail to recover dense lines in ``img\_042'' and ``img\_093'', while their MS$^3$-Conv counterparts correctly restore the lattice pattern. For image ``img\_098'', we observe that SRResNet cannot recover the lattices and CARN would suffer from blurring effects. In contrast, their MS$^3$-Conv versions can alleviate the blurring effects and recover more details.
However, when focusing on flat regions and smooth edges (see Figure \ref{fig:comp_smooth}), we can observe a large gap on PSNR comparing multi-scale with standard convolutions. Counterintuitively, this does not necessarily indicate a significant perceptual difference. For image ``0830'', the PSNR of SRResNet with MS$^3$-Conv decreases $3.4$ dB over SRResNet, but the two images are perceptually indistinguishable. The region-based variation implies that these networks have different preferences to different structures.





To further illustrate the above analyses, we show visual comparisons for multiple scales of images in Figure \ref{fig:ms-dataset}. We change the scale factor $\{0.8, 0.6, 0.5\}$ and rescale each image to explore how these networks behave differently on recovering patterns. Figure \ref{fig:ms-dataset} shows that SRResNet with MS$^3$-Conv performs consistently better than its standard convolution counterpart at all scales (see yellow box). We also notice that both standard convolution and MS$^3$-Conv on CARN backbone can correctly recover the structure of building with scaling factor $0.8$ and $0.6$. However, when using a lower scale ($0.5$), the results by standard convolution would lose the structures and exhibit aliasing effects (see blue box). In contrast, MS$^3$-Conv-equipped CARN can alleviate this problem and recover the right pattern.

\section{Conclusion}
In this paper, we present a unified formulation over various multi-scale structures. Under this framework, we provide a comprehensive investigation on variants of multi-scale convolution. Based on the investigation, we propose a generic and efficient multi-scale convolution unit -- Multi-Scale cross-Scale Share-weights convolution (MS$^3$-Conv).
Our results indicate that the proposed MS$^3$-Conv can achieve better performance than the standard convolution with less parameters and computational cost. 
We also comprehensively study the visual quality, which show that MS$^3$-Conv behave better to recover high-frequency details.

\textbf{Acknowledgements.} This work is partially supported by National Key Research and Development Program of China (2016YFC1400704), Shenzhen Research Program (JCYJ20170818164704758, JCYJ20150925163005055, \\CXB201104220032A), and Joint Lab of CAS-HK.

%
%
\bibliographystyle{splncs04}
\bibliography{egbib}

\begin{thebibliography}{10}
\providecommand{\url}[1]{\texttt{#1}}
\providecommand{\urlprefix}{URL }
\providecommand{\doi}[1]{https://doi.org/#1}

\bibitem{Agustsson_2017_CVPR_Workshops}
Agustsson, E., Timofte, R.: Ntire 2017 challenge on single image
  super-resolution: Dataset and study. In: The IEEE Conference on Computer
  Vision and Pattern Recognition (CVPR) Workshops (July 2017)

\bibitem{ahn2018fast}
Ahn, N., Kang, B., Sohn, K.A.: Fast, accurate, and lightweight super-resolution
  with cascading residual network. In: Proceedings of the European Conference
  on Computer Vision (ECCV). pp. 252--268 (2018)

\bibitem{bevilacqua2012low}
Bevilacqua, M., Roumy, A., Guillemot, C., Alberi-Morel, M.L.: Low-complexity
  single-image super-resolution based on nonnegative neighbor embedding  (2012)

\bibitem{chen2018big}
Chen, C.F., Fan, Q., Mallinar, N., Sercu, T., Feris, R.: Big-little net: An
  efficient multi-scale feature representation for visual and speech
  recognition. arXiv preprint arXiv:1807.03848  (2018)

\bibitem{chen2019drop}
Chen, Y., Fang, H., Xu, B., Yan, Z., Kalantidis, Y., Rohrbach, M., Yan, S.,
  Feng, J.: Drop an octave: Reducing spatial redundancy in convolutional neural
  networks with octave convolution. arXiv preprint arXiv:1904.05049  (2019)

\bibitem{dong2014learning}
Dong, C., Loy, C.C., He, K., Tang, X.: Learning a deep convolutional network
  for image super-resolution. In: European conference on computer vision. pp.
  184--199. Springer (2014)

\bibitem{dong2016image}
Dong, C., Loy, C.C., He, K., Tang, X.: Image super-resolution using deep
  convolutional networks. IEEE transactions on pattern analysis and machine
  intelligence  \textbf{38}(2),  295--307 (2016)

\bibitem{dong2016accelerating}
Dong, C., Loy, C.C., Tang, X.: Accelerating the super-resolution convolutional
  neural network. In: European Conference on Computer Vision. pp. 391--407.
  Springer (2016)

\bibitem{gu2019blind}
Gu, J., Lu, H., Zuo, W., Dong, C.: Blind super-resolution with iterative kernel
  correction. In: Proceedings of the IEEE Conference on Computer Vision and
  Pattern Recognition. pp. 1604--1613 (2019)

\bibitem{he2016identity}
He, K., Zhang, X., Ren, S., Sun, J.: Identity mappings in deep residual
  networks. In: European conference on computer vision. pp. 630--645. Springer
  (2016)

\bibitem{huang2017multi}
Huang, G., Chen, D., Li, T., Wu, F., van~der Maaten, L., Weinberger, K.Q.:
  Multi-scale dense networks for resource efficient image classification. arXiv
  preprint arXiv:1703.09844  (2017)

\bibitem{huang2015single}
Huang, J.B., Singh, A., Ahuja, N.: Single image super-resolution from
  transformed self-exemplars. In: Proceedings of the IEEE Conference on
  Computer Vision and Pattern Recognition. pp. 5197--5206 (2015)

\bibitem{ke2017multigrid}
Ke, T.W., Maire, M., Yu, S.X.: Multigrid neural architectures. In: Proceedings
  of the IEEE Conference on Computer Vision and Pattern Recognition. pp.
  6665--6673 (2017)

\bibitem{kim2016accurate}
Kim, J., Kwon~Lee, J., Mu~Lee, K.: Accurate image super-resolution using very
  deep convolutional networks. In: Proceedings of the IEEE conference on
  computer vision and pattern recognition. pp. 1646--1654 (2016)

\bibitem{kingma2014adam}
Kingma, D.P., Ba, J.: Adam: A method for stochastic optimization. CoRR
  \textbf{abs/1412.6980} (2015)

\bibitem{lai2017deep}
Lai, W.S., Huang, J.B., Ahuja, N., Yang, M.H.: Deep laplacian pyramid networks
  for fast and accurate superresolution. In: IEEE Conference on Computer Vision
  and Pattern Recognition. vol.~2, p.~5 (2017)

\bibitem{ledig2017photo}
Ledig, C., Theis, L., Husz{\'a}r, F., Caballero, J., Cunningham, A., Acosta,
  A., Aitken, A.P., Tejani, A., Totz, J., Wang, Z., et~al.: Photo-realistic
  single image super-resolution using a generative adversarial network. In:
  CVPR. vol.~2, p.~4 (2017)

\bibitem{li2019scale}
Li, Y., Chen, Y., Wang, N., Zhang, Z.: Scale-aware trident networks for object
  detection. arXiv preprint arXiv:1901.01892  (2019)

\bibitem{lim2017enhanced}
Lim, B., Son, S., Kim, H., Nah, S., Lee, K.M.: Enhanced deep residual networks
  for single image super-resolution. In: The IEEE conference on computer vision
  and pattern recognition (CVPR) workshops. vol.~1, p.~4 (2017)

\bibitem{lin2017feature}
Lin, T.Y., Doll{\'a}r, P., Girshick, R., He, K., Hariharan, B., Belongie, S.:
  Feature pyramid networks for object detection. In: Proceedings of the IEEE
  conference on computer vision and pattern recognition. pp. 2117--2125 (2017)

\bibitem{martin2001database}
Martin, D., Fowlkes, C., Tal, D., Malik, J.: A database of human segmented
  natural images and its application to evaluating segmentation algorithms and
  measuring ecological statistics. In: Computer Vision, 2001. ICCV 2001.
  Proceedings. Eighth IEEE International Conference on. vol.~2, pp. 416--423.
  IEEE (2001)

\bibitem{newell2016stacked}
Newell, A., Yang, K., Deng, J.: Stacked hourglass networks for human pose
  estimation. In: European conference on computer vision. pp. 483--499.
  Springer (2016)

\bibitem{2017-Paszke-p-}
Paszke, A., Gross, S., Chintala, S., Chanan, G., Yang, E., DeVito, Z., Lin, Z.,
  Desmaison, A., Antiga, L., Lerer, A.: Automatic differentiation in pytorch
  (2017)

\bibitem{ronneberger2015u}
Ronneberger, O., Fischer, P., Brox, T.: U-net: Convolutional networks for
  biomedical image segmentation. In: International Conference on Medical image
  computing and computer-assisted intervention. pp. 234--241. Springer (2015)

\bibitem{shi2016real}
Shi, W., Caballero, J., Husz{\'a}r, F., Totz, J., Aitken, A.P., Bishop, R.,
  Rueckert, D., Wang, Z.: Real-time single image and video super-resolution
  using an efficient sub-pixel convolutional neural network. In: Proceedings of
  the IEEE Conference on Computer Vision and Pattern Recognition. pp.
  1874--1883 (2016)

\bibitem{sun2019high}
Sun, K., Zhao, Y., Jiang, B., Cheng, T., Xiao, B., Liu, D., Mu, Y., Wang, X.,
  Liu, W., Wang, J.: High-resolution representations for labeling pixels and
  regions. arXiv preprint arXiv:1904.04514  (2019)

\bibitem{tong2017image}
Tong, T., Li, G., Liu, X., Gao, Q.: Image super-resolution using dense skip
  connections. In: Computer Vision (ICCV), 2017 IEEE International Conference
  on. pp. 4809--4817. IEEE (2017)

\bibitem{wang2018elastic}
Wang, H., Kembhavi, A., Farhadi, A., Yuille, A., Rastegari, M.: Elastic:
  Improving cnns with instance specific scaling policies. arXiv preprint
  arXiv:1812.05262  (2018)

\bibitem{wang2018recovering}
Wang, X., Yu, K., Dong, C., Change~Loy, C.: Recovering realistic texture in
  image super-resolution by deep spatial feature transform. In: Proceedings of
  the IEEE Conference on Computer Vision and Pattern Recognition. pp. 606--615
  (2018)

\bibitem{wang2018esrgan}
Wang, X., Yu, K., Wu, S., Gu, J., Liu, Y., Dong, C., Qiao, Y., Change~Loy, C.:
  Esrgan: Enhanced super-resolution generative adversarial networks. In:
  Proceedings of the European Conference on Computer Vision (ECCV). pp.~0--0
  (2018)

\bibitem{yang2010image}
Yang, J., Wright, J., Huang, T.S., Ma, Y.: Image super-resolution via sparse
  representation. IEEE transactions on image processing  \textbf{19}(11),
  2861--2873 (2010)

\bibitem{yu2015multi}
Yu, F., Koltun, V.: Multi-scale context aggregation by dilated convolutions.
  arXiv preprint arXiv:1511.07122  (2015)

\bibitem{zhang2018image}
Zhang, Y., Li, K., Li, K., Wang, L., Zhong, B., Fu, Y.: Image super-resolution
  using very deep residual channel attention networks. In: Proceedings of the
  European Conference on Computer Vision (ECCV). pp. 286--301 (2018)

\bibitem{zhang2018residual}
Zhang, Y., Tian, Y., Kong, Y., Zhong, B., Fu, Y.: Residual dense network for
  image super-resolution. In: The IEEE Conference on Computer Vision and
  Pattern Recognition (CVPR) (2018)

\bibitem{zhao2017pyramid}
Zhao, H., Shi, J., Qi, X., Wang, X., Jia, J.: Pyramid scene parsing network.
  In: Proceedings of the IEEE conference on computer vision and pattern
  recognition. pp. 2881--2890 (2017)

\end{thebibliography}


\begin{thebibliography}{10}
\providecommand{\url}[1]{\texttt{#1}}
\providecommand{\urlprefix}{URL }
\providecommand{\doi}[1]{https://doi.org/#1}

\bibitem{ahn2018fast}
Ahn, N., Kang, B., Sohn, K.A.: Fast, accurate, and lightweight super-resolution
  with cascading residual network. In: Proceedings of the European Conference
  on Computer Vision (ECCV). pp. 252--268 (2018)

\bibitem{dong2016image}
Dong, C., Loy, C.C., He, K., Tang, X.: Image super-resolution using deep
  convolutional networks. IEEE transactions on pattern analysis and machine
  intelligence  \textbf{38}(2),  295--307 (2016)

\bibitem{dong2016accelerating}
Dong, C., Loy, C.C., Tang, X.: Accelerating the super-resolution convolutional
  neural network. In: European Conference on Computer Vision. pp. 391--407.
  Springer (2016)

\bibitem{kim2016accurate}
Kim, J., Kwon~Lee, J., Mu~Lee, K.: Accurate image super-resolution using very
  deep convolutional networks. In: Proceedings of the IEEE conference on
  computer vision and pattern recognition. pp. 1646--1654 (2016)

\bibitem{lai2017deep}
Lai, W.S., Huang, J.B., Ahuja, N., Yang, M.H.: Deep laplacian pyramid networks
  for fast and accurate superresolution. In: IEEE Conference on Computer Vision
  and Pattern Recognition. vol.~2, p.~5 (2017)

\bibitem{lim2017enhanced}
Lim, B., Son, S., Kim, H., Nah, S., Lee, K.M.: Enhanced deep residual networks
  for single image super-resolution. In: The IEEE conference on computer vision
  and pattern recognition (CVPR) workshops. vol.~1, p.~4 (2017)

\bibitem{tai2017image}
Tai, Y., Yang, J., Liu, X.: Image super-resolution via deep recursive residual
  network. In: Proceedings of the IEEE Conference on Computer Vision and
  Pattern Recognition. vol.~1, p.~5 (2017)

\bibitem{tai2017memnet}
Tai, Y., Yang, J., Liu, X., Xu, C.: Memnet: A persistent memory network for
  image restoration. In: Proceedings of the IEEE Conference on Computer Vision
  and Pattern Recognition. pp. 4539--4547 (2017)

\bibitem{tong2017image}
Tong, T., Li, G., Liu, X., Gao, Q.: Image super-resolution using dense skip
  connections. In: Computer Vision (ICCV), 2017 IEEE International Conference
  on. pp. 4809--4817. IEEE (2017)

\bibitem{wang2018esrgan}
Wang, X., Yu, K., Wu, S., Gu, J., Liu, Y., Dong, C., Qiao, Y., Change~Loy, C.:
  Esrgan: Enhanced super-resolution generative adversarial networks. In:
  Proceedings of the European Conference on Computer Vision (ECCV). pp.~0--0
  (2018)

\bibitem{zhang2018learning}
Zhang, K., Zuo, W., Zhang, L.: Learning a single convolutional super-resolution
  network for multiple degradations. In: IEEE Conference on Computer Vision and
  Pattern Recognition. vol.~6 (2018)

\bibitem{zhang2018image}
Zhang, Y., Li, K., Li, K., Wang, L., Zhong, B., Fu, Y.: Image super-resolution
  using very deep residual channel attention networks. In: Proceedings of the
  European Conference on Computer Vision (ECCV). pp. 286--301 (2018)

\bibitem{zhang2018residual}
Zhang, Y., Tian, Y., Kong, Y., Zhong, B., Fu, Y.: Residual dense network for
  image super-resolution. In: The IEEE Conference on Computer Vision and
  Pattern Recognition (CVPR) (2018)

\end{thebibliography}
\end{document}


\pagestyle{headings}
\mainmatter
\def\ECCVSubNumber{3269}  

\title{Exploring Multi-Scale Feature Propagation and Communication for Image Super Resolution\\Supplementary Materials} 


\titlerunning{Multi-Scale Feature Propagation and Communication}
%
\author{
Ruicheng Feng\inst{1} \and
Weipeng Guan\inst{1} \and
Yu qiao\inst{1,2} \and
Chao Dong\inst{1}
}
\authorrunning{R. Feng et al.}
%
\institute{\inst{1}Shenzhen Key Lab of Computer Vision and Pattern Recognition, SIAT-SenseTime Joint Lab, Shenzhen Institutes of Advanced Technology, Chinese Acedamy of Sciences \\
\inst{2}The Chinese University of Hong Kong\\
\email{\{rc.feng, wp.guan, yu.qiao, chao.dong\}@siat.ac.cn}}
\maketitle

\begin{abstract}
In this supplementary file, we first show the comparison with other state-of-the-art methods in Section \ref{comparison}. We also demonstrate that our MS$^3$-Conv-based models can outperform all state-of-the-art models that have less than $5$M parameters with least parameters and computation cost. In Section \ref{visual}, we present more qualitative results of multi-scale  convolution for visual comparison.
\end{abstract}

\section{Comparison with State-of-the-art Methods}
\label{comparison}
In this section, we compare the proposed MS$^3$-Conv with state-of-the-art SR methods on PSNR and SSIM metrics.  Table \ref{table:comparison} shows the quantitative comparisons. Note that we group several large models into another part, as they all have a considerable amount of parameters (e.g., $16$M for RCAN). It is not surprising that EDSR \cite{lim2017enhanced}, RDN \cite{zhang2018residual}, RCAN \cite{zhang2018image}, RRDB \cite{wang2018esrgan} (see the lower part) achieve improvements over ours, because they have $43$M, $13$M, $15$M, $16$M parameters, respectively, which are nearly $33\times$ and roughly $10\times$ more than our CARN-MS$^3$-Conv$+$.

It demonstrates that our CARN-MS$^3$-Conv$+$ outperforms other models that have less than $2$M parameters (the upper part). In particular, comparing models that have roughly similar number of parameters, e.g., SRMDNF \cite{zhang2018learning}, SRDenseNet \cite{tong2017image}, CARN \cite{ahn2018fast}, CARN-MS$^3$-Conv$+$ obtains peak performance, while achieving roughly $5\times$ speedup over SRMDNF and SRDenseNet. Also, this model achieves better performance than the original CARN with less FLOPs, which validates the effectiveness of the modification. Note that our implemented CARN backbone network has $35.5$ GFLOPs as a redundant convolution layer added at the end of the network, which attributes around $10$ GFLOPs to the network but brings minor improvement. Compared to this backbone network, our CARN-MS$^3$-Conv$+$ saves $40\%$ computation cost.

Moreover, CARN-MS$^3$-Conv shows comparable results against computationally-expensive models (e.g., SRDenseNet), while only requiring the similar amount of FLOPs with respect to SRCNN ($15.06$ vs. $14.99$ GFLOPs). This significant improvement can be mainly attributed to two fold: 1) the cascading mechanism and the network architecture proposed by CARN. 2) Our MS$^3$-Conv modification.

\begin{figure*}[ht]
\centering
\includegraphics[width=0.98\linewidth]{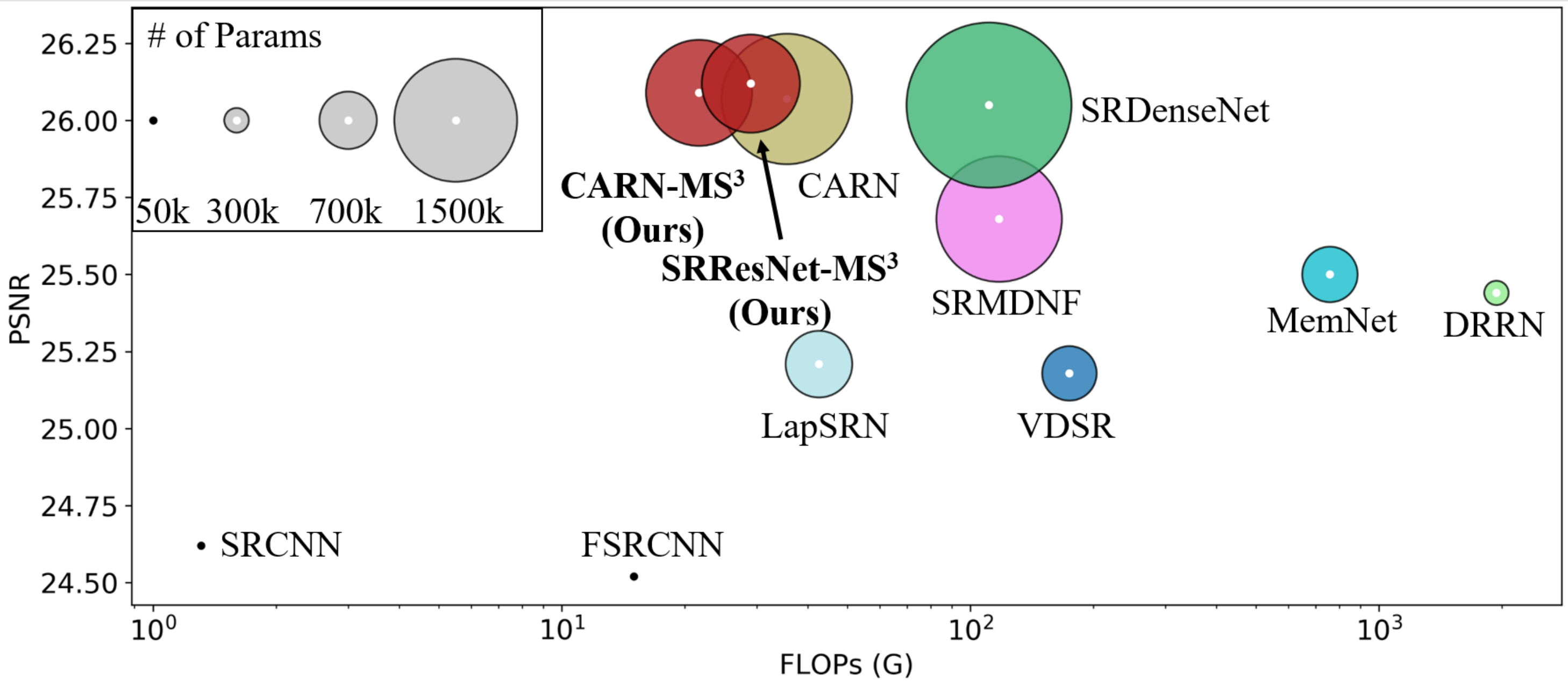}
\caption{Trade-off between performance vs. FLOPs and parameters
on Urban100 ($\times4$) dataset. The size of the circle represents the number of parameters.}
\label{fig:bubble}
\end{figure*}

In Figure \ref{fig:bubble}, we compare our MS$^3$-Conv on different backbone networks against the various benchmark algorithms in terms of the FLOPs and the number of parameters on the Urban100 ($\times4$) dataset. The figure shown that our MS$^3$-Conv-based models can outperform all state-of-the-art models that have less than 5M parameters with least parameters and computation cost. Especially, CARN-MS$^3$-Conv and SRResNet-MS$^3$-Conv obtain comparable or even better results with CARN and SRDenseNet, but require much less parameters and FLOPs.

\begin{table*}[t]
\begin{center}
\caption{Quantitative evaluation (PSNR) for scaling factor $4$ on benchmark dataset Set5, Set14, BSD100, Urban100. The best two results in each part are highlighted in {\color{red}red} and {\color{blue}blue} colors, respectively.}
\vskip -0.2cm
\label{table:comparison}
\begin{tabular}{lrrcccc}
\hline\hline\noalign{\smallskip}
\multirow{2}{*}{Model} & \multirow{2}{*}{\# Params (K)} & \multirow{2}{*}{FLOPs (G)} & Set5 & Set14 & BSD100 & Urban100\\
&&& PSNR / SSIM & PSNR / SSIM & PSNR / SSIM & PSNR / SSIM\\
\noalign{\smallskip}
\hline
SRCNN \cite{dong2016image} 		& $57$ 		& $14.99$ 	& $30.48 / 0.8628$ & $27.49 / 0.7503$ & $26.90 / 0.7101$ & $24.52 / 0.7221$\\
FSRCNN \cite{dong2016accelerating} 		& $12$ 		& $1.31$ 		& $30.71 / 0.8657$ & $27.59 / 0.7535$ & $26.98 / 0.7150$ & $24.62 / 0.7280$\\
VDSR \cite{kim2016accurate} 		& $665$ 		& $174.25$ 	& $31.35 / 0.8838$ & $28.01 / 0.7674$ & $27.29 / 0.7251$ & $25.18 / 0.7524$\\
LapSRN \cite{lai2017deep} 		& $813$ 		& $42.50$ 	& $31.54 / 0.8850$ & $28.19 / 0.7720$ & $27.32 / 0.7280$ & $25.21 / 0.7560$\\
DRRN \cite{tai2017image} 		& $297$ 		& $1,933.34$ 	& $31.68 / 0.8888$ & $28.21 / 0.7720$ & $27.38 / 0.7284$ & $25.44 / 0.7638$\\
MemNet \cite{tai2017memnet} 		& $677$ 		& $757.30$ 	& $31.74 / 0.8893$ & $28.26 / 0.7723$ & $27.40 / 0.7281$ & $25.50 / 0.7630$\\
SRMDNF \cite{zhang2018learning} & $1,533$ & $117.42$ & $31.96 / 0.8925$ & $28.35 / 0.7787$ & $27.49 / 0.7337$ & $25.68 / 0.7731$ \\
SRDenseNet \cite{tong2017image} 	& $2,015$ 	& $110.90$ 	& $32.02 / 0.8934$ & $28.50 / 0.7782$ & $27.53 / 0.7337$ & $26.05 / 0.7819$\\
CARN-MS$^3$-Conv & $450$ & $15.06$ & $32.02 / 0.8931$ & $28.50 / 0.7795$ & $27.51 / 0.7339$ & $25.81 / 0.7765$\\
CARN \cite{ahn2018fast} 		& $1,592$ 	& $25.86$ 	& {\color{red}$32.13$}$/${\color{blue}$0.8937$} & {\color{blue}$28.60 / 0.7806$} & {\color{blue}$27.58 / 0.7349$} & {\color{blue}$26.07 / 0.7837$}\\
CARN-MS$^3$-Conv+ & $1,300$ & $21.63$ & {\color{blue}$32.09$}$/${\color{red}$0.8945$} & {\color{red}$28.61 / 0.7821$} & {\color{red}$27.59 / 0.7368$} &  {\color{red}$26.09 / 0.7859$}\\
\hline\hline
EDSR \cite{lim2017enhanced} 	& $43,090$ 	&  $823.32$ & $32.46 / 0.8968$ & $28.80 / 0.7876$ & $27.71 / 0.7420$ & $26.64 / 0.8033$ \\
RDN \cite{zhang2018residual} & $12,834$ & $226.91$ & $32.47 / 0.8990$ & $28.81/0.7871$ & $27.72/0.7419$ & $26.61/0.8028$ \\
RCAN \cite{zhang2018image} & $15,322$ & $248.72$ & {\color{red}$32.63 / 0.9002$} & {\color{blue}$28.87/0.7889$} & {\color{red}$27.77/0.7436$} & {\color{red}$26.82/0.8087$} \\
RRDB \cite{wang2018esrgan} & $16,919$ & $293.71$ & {\color{blue}$32.60 / 0.9002$} & {\color{red}$28.88/0.7896$} & {\color{blue}$27.76/0.7432$} & {\color{blue}$26.73/0.8072$} \\
\hline\hline
\end{tabular}
\end{center}
\end{table*}

\section{More qualitative comparison}
\label{visual}
\begin{figure*}[ht]
\centering
\begin{subfigure}[b]{0.99\linewidth}
\centering
\includegraphics[width=0.9\linewidth]{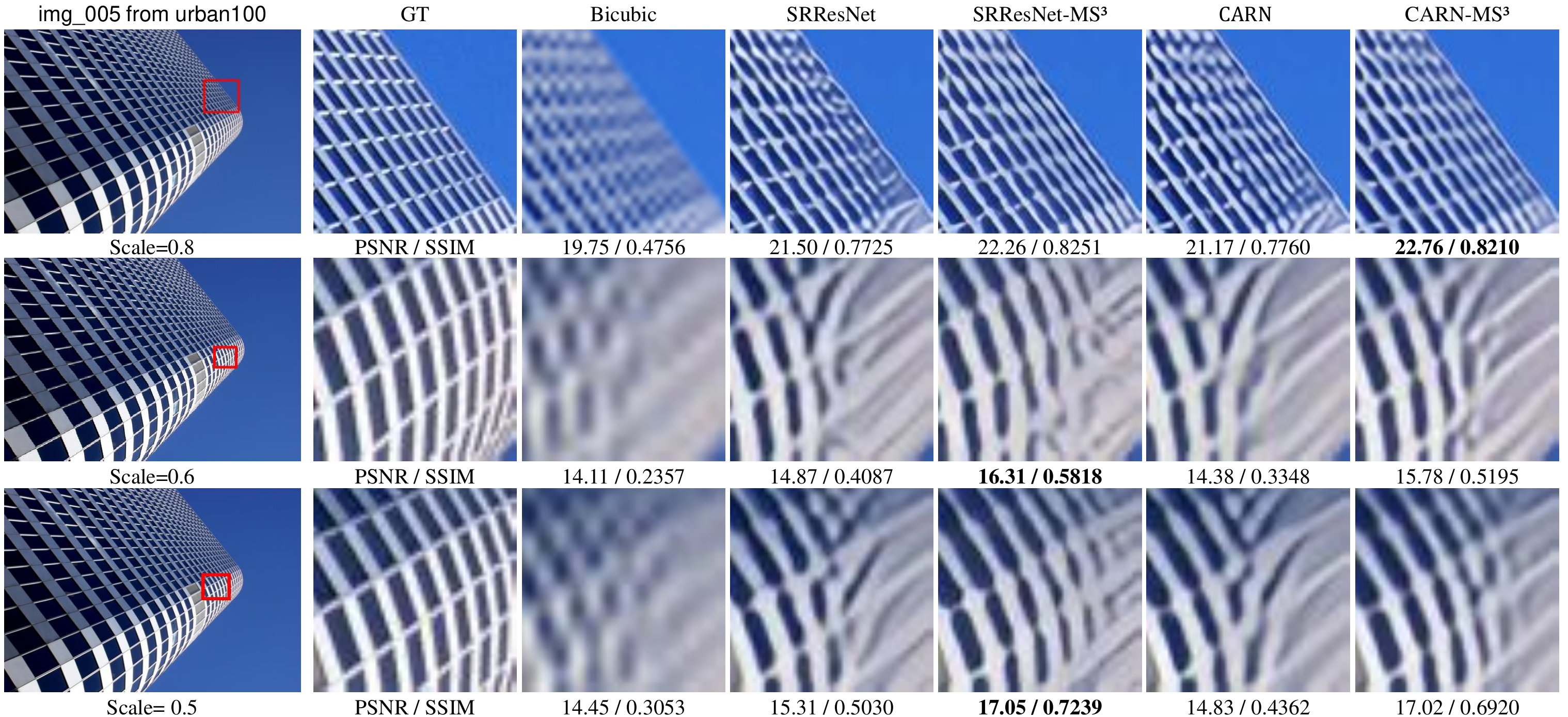}
\end{subfigure}
\begin{subfigure}[b]{0.99\linewidth}
\centering
\includegraphics[width=0.9\linewidth]{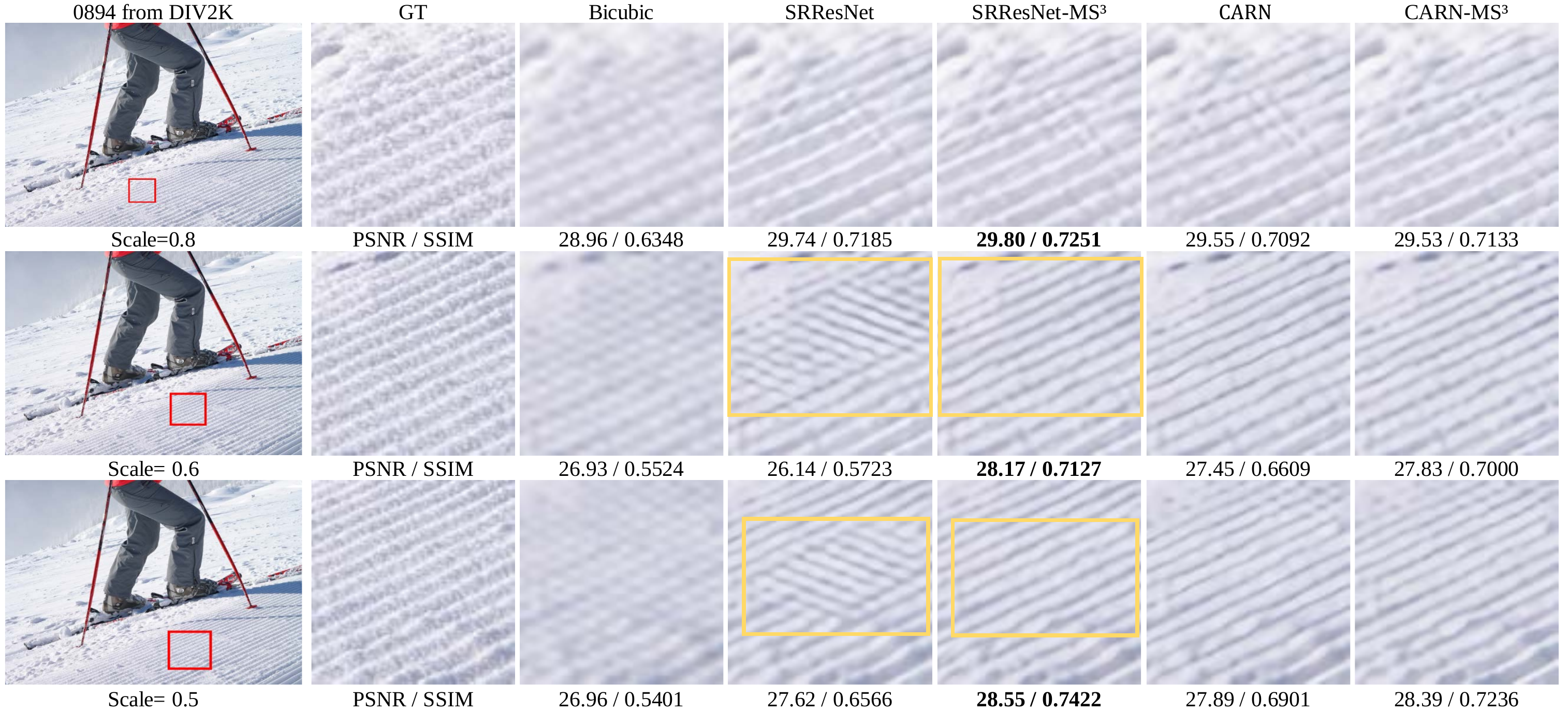}
\end{subfigure}

\begin{subfigure}[b]{0.99\linewidth}
\centering
\includegraphics[width=0.9\linewidth]{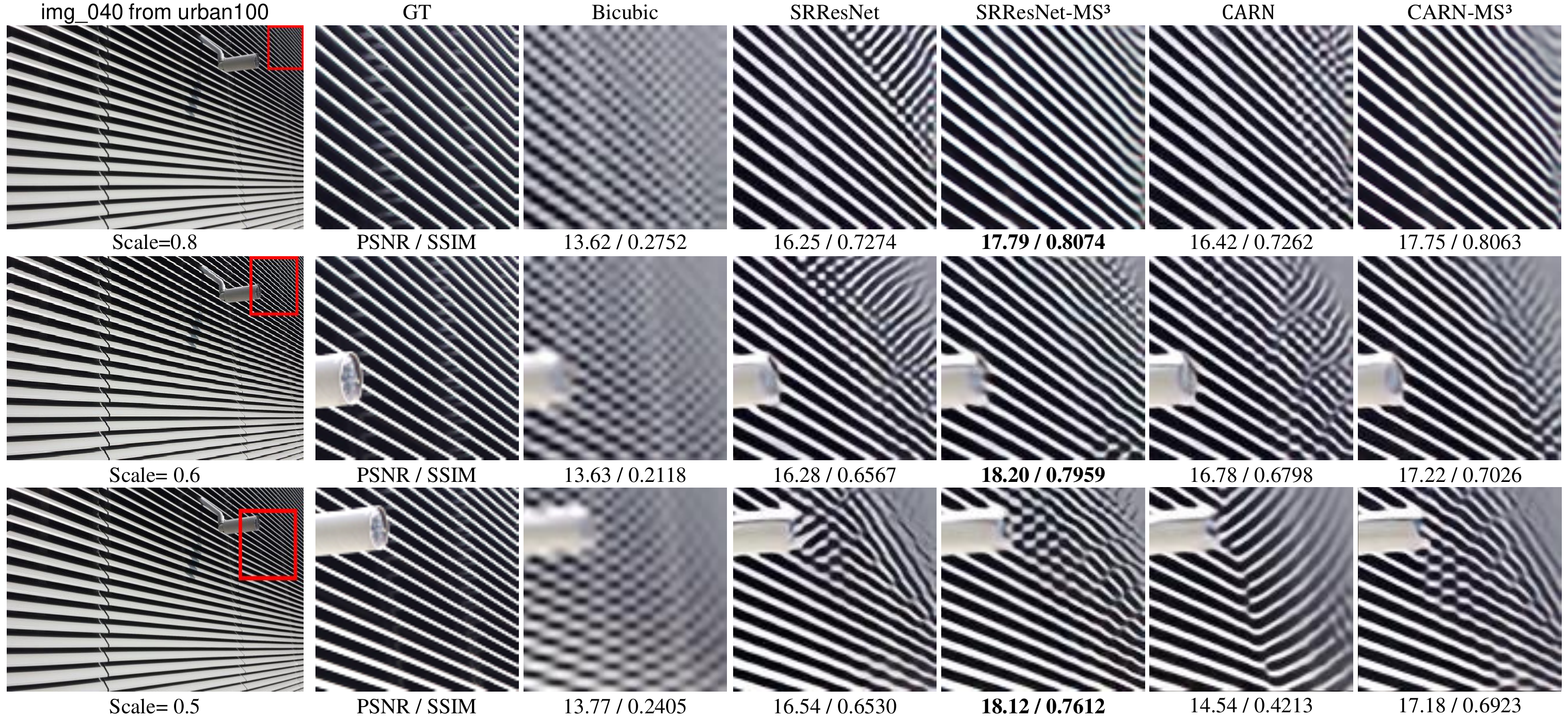}
\end{subfigure}
\caption{Qualitative comparisons on rescaling images, with scaling factor {$0.8$, $0.6$, $0.5$}. All images are rescaled to the same resolution for better visualization. \textbf{Regions of interest in yellow boxes.}}
\label{fig:visual1}
\vskip -0.3cm
\end{figure*}

\begin{figure*}[ht]
\centering
\begin{subfigure}[b]{0.99\linewidth}
\centering
\includegraphics[width=0.9\linewidth]{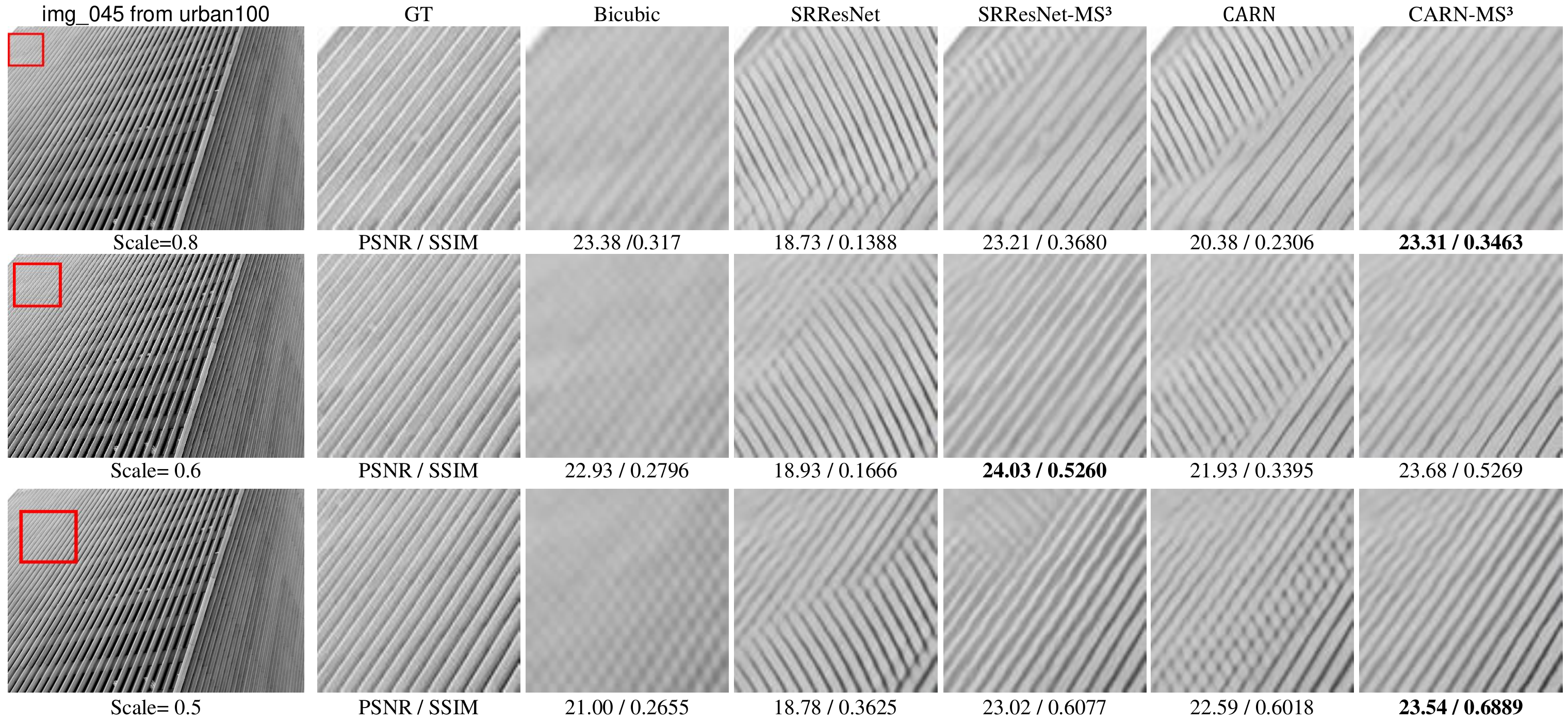}
\end{subfigure}

\begin{subfigure}[b]{0.99\linewidth}
\centering
\includegraphics[width=0.9\linewidth]{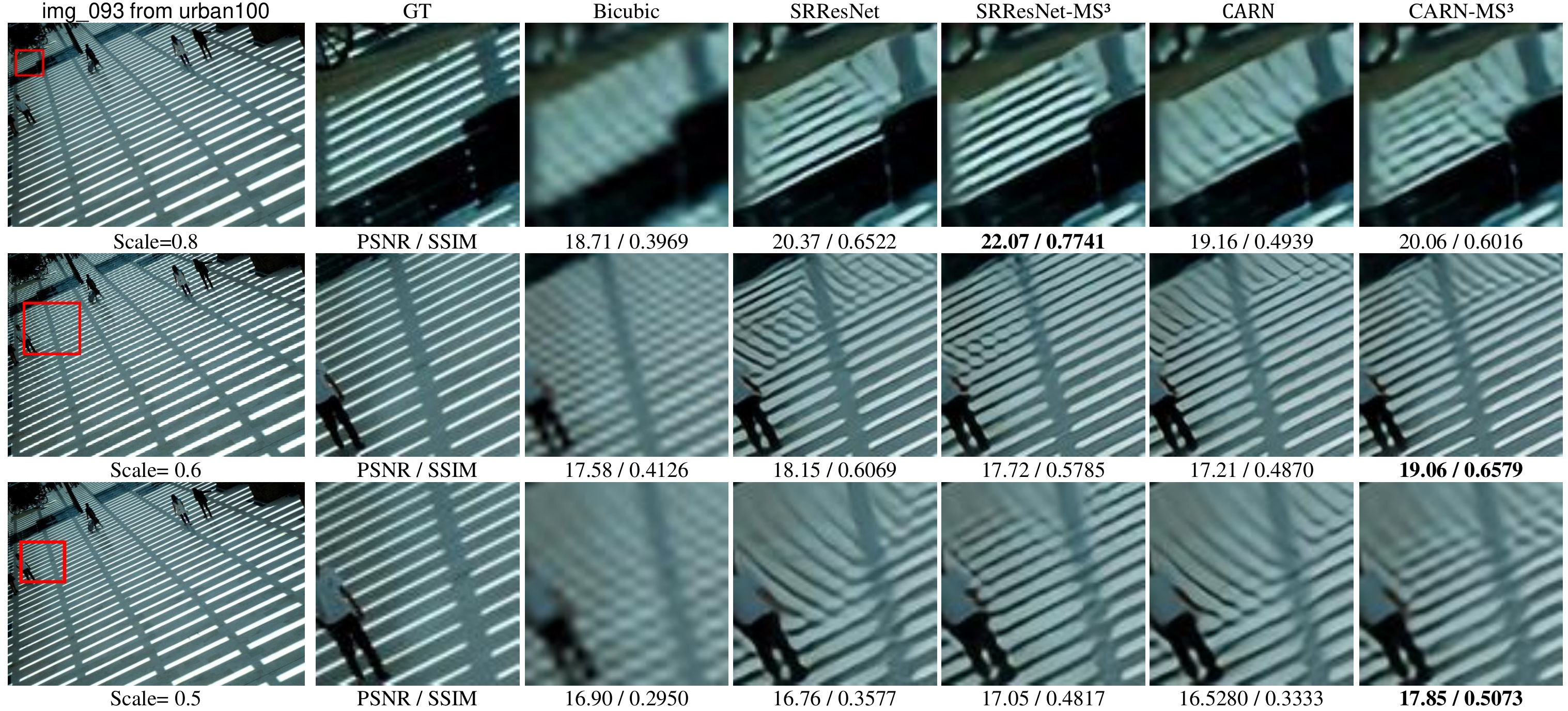}
\end{subfigure}

\begin{subfigure}[b]{0.99\linewidth}
\centering
\includegraphics[width=0.9\linewidth]{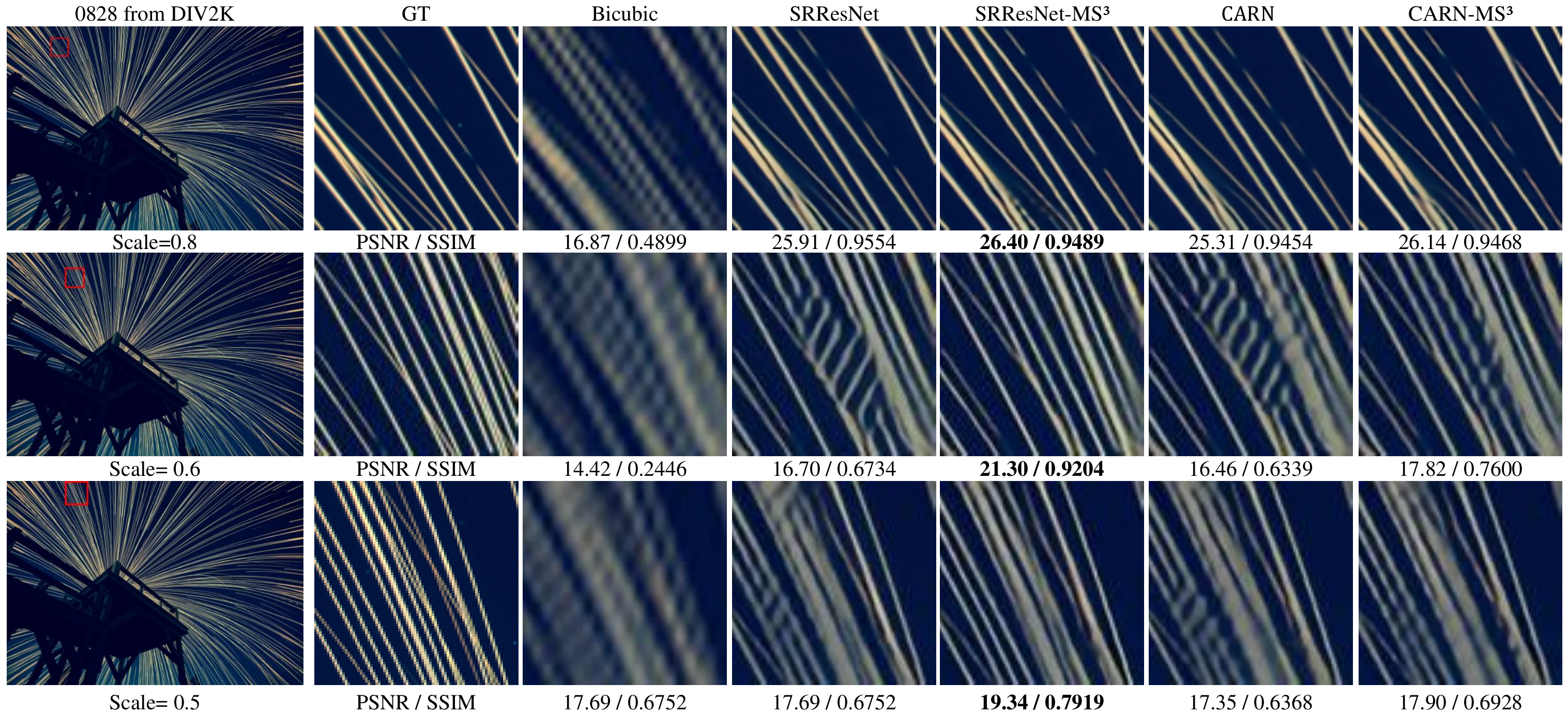}
\end{subfigure}

\caption{Qualitative comparisons on rescaling images, with scaling factor {$0.8$, $0.6$, $0.5$}. All images are rescaled to the same resolution for better visualization.}
\label{fig:visual2}
\vskip -0.3cm
\end{figure*}

To illustrate the qualitative analyses, we show visual comparisons for multiple scales of images in Figure \ref{fig:visual1} and Figure \ref{fig:visual2}. We change the scale factor $\{0.8, 0.6, 0.5\}$ and rescale each image to explore how these networks behave differently on recovering patterns. The image ``img\_005'' and ``img\_040'' show that SRResNet with MS$^3$-Conv performs consistently better than its standard convolution counterpart at all scales.
Most of the results reveal that MS$^3$-Conv on different backbone networks tend to correctly recover the right lattice pattern, while the results by standard convolution would lose the structures, which demonstrates the effectiveness of multi-scale convolutions.

\clearpage
\bibliographystyle{splncs04}
\bibliography{egbib}